\documentclass{article}
\usepackage{amssymb,amsmath}
\usepackage{graphicx}
\usepackage{booktabs}
\usepackage[official]{eurosym}
\usepackage{subfig}
\usepackage[usenames,dvipsnames,table]{xcolor}
\usepackage[T1]{fontenc}
\usepackage[utf8]{inputenc}
\usepackage{authblk}
\newcommand{\HRule}{\rule{\linewidth}{0.5mm}}

\title{Proposal for SPS beam time for the baby MIND and TASD neutrino detector prototypes}
\author[a]{R. Asfandiyarov}
\author[b]{R. Bayes}
\author[a]{A. Blondel}
\author[c]{M. Bogomilov}
\author[d]{A. Bross}
\author[a]{F. Cadoux}
\author[e]{A. Cervera}
\author[f]{A. Izmaylov}
\author[a]{Y. Karadzhov}
\author[f]{I. Karpikov}
\author[f]{M. Khabibulin}
\author[f]{A. Khotyantsev}
\author[f]{A. Kopylov}
\author[f]{Y. Kudenko}
\author[c]{R. Matev}
\author[f]{O. Mineev}
\author[f]{Y. Musienko}
\author[g]{M. Nessi}
\author[a]{E. Noah}
\author[h]{A. Rubbia}
\author[f]{A. Shaykiev}
\author[b]{P. Soler}
\author[c]{R. Tsenov}
\author[c]{G. Vankova-Kirilova}
\author[f]{N. Yershov}

\affil[a]{University of Geneva, Section de Physique, DPNC, Geneva, Switzerland}
\affil[b]{University of Glasgow, School of Physics and Astronomy, Glasgow, UK}
\affil[c]{University of Sofia, Department of Physics, Sofia, Bulgaria}
\affil[d]{Fermi National Accelerator Laboratory, Batavia, Illinois, USA}
\affil[e]{IFIC (CSIC \& University of Valencia), Valencia, Spain}
\affil[f]{Institute of Nuclear Research, Russian Academy of Sciences, Moscow, Russia}
\affil[g]{European Organization for Nuclear Research, CERN, Geneva, Switzerland}
\affil[h]{ETH Zurich, Institute for Particle Physics, Zurich, Switzerland}

\begin{document}
\maketitle
%Title page
\begin{titlepage}
\begin{center}
%\includegraphics[width=0.15\textwidth]
%{figs-lbno-ND-proto/fig-neutrino-group-logo}\\[1cm]
%\textsc{\LARGE University of Geneva}\\[1.5cm]
%\textsc{\LARGE Technical Note}\\[0.5cm]
%  TITLE
\HRule \\[0.4cm]
{\huge \bfseries Proposal for SPS beam time for the baby MIND and TASD neutrino detector prototypes}\\[0.4cm]
{\huge \bfseries - v1.0 - }\\[0.4cm]
\HRule \\[1.5cm]
% Author details
\begin{minipage}{0.4\textwidth}
\begin{flushleft} \large
\emph{Testbeam contact:}\\
Etam \textsc{Noah}
\end{flushleft}
\end{minipage}
%\begin{minipage}{0.4\textwidth}
%\begin{flushright} \large
%\emph{AIDA Task Proposers:} \\
%Alain \textsc{Blondel}\\
%Paul \textsc{Soler}
%\end{flushright}
%\end{minipage}

\textsc{}\\[1.5cm]
\textsc{On behalf of collaborating partners}\\[1.5cm]

\section*{Abstract}
The design, construction and testing of neutrino detector prototypes at CERN are ongoing activities. This document reports on the design of solid state baby MIND and TASD detector prototypes and outlines requirements for a test beam at CERN to test these, tentatively planned on the H8 beamline in the North Area, which is equipped with a large aperture magnet. It is hoped that this will allow for the current proposal to be considered in light of the recently approved projects related to neutrino activities with the SPS in the North Area in the medium term 2015-2020.

\vfill
\date{May 23, 2014}
%{\large \today}
\end{center}
\end{titlepage}

\tableofcontents
\pagestyle{plain}
\mbox{}

\clearpage

\section{Introduction}

Operational neutrino facilities and those about to go online in the next decade such as accelerator-based experiments NO$\nu$A, T2K, MINOS, ICARUS and OPERA and reactor-based experiments Double Chooz, Daya Bay and Reno are expected to provide increasing precision on measured parameters such as $\theta_{13}$, $\theta_{23}$, $\vert$$\Delta$$m^2_{31}$$\vert$.
A re-assessment is underway as to how to study the established remaining unknowns such as the existence of CP-violation, its phase if it does exist, $\delta$CP, neutrino mass hierarchy and the postulated existence of sterile neutrinos with next-generation neutrino facilities such as the planned accelerator-based facilities:
\begin{itemize}
\item Long baseline neutrino oscillation experiment (LBNO)\footnotemark, Expression of Interest (EoI) CERN-SPSC-2012-021 (SPSC-EOI-007) \cite{Stahl:2012exa}. 
\item A prototype muon storage ring proposal, $\nu$STORM (LOI May 2012) \cite{Kyberd:2012iz}.
\item The Hyper-Kamiokande Experiment in Japan (LOI September 2011) \cite{Abe:2011ts}.
\item Short baseline neutrino experiment, CERN-SPSC-2012-010 (SPSC-P-347) \cite{Antonello:2012hf}.
\item The long baseline neutrino experiment (LBNE) in the USA \cite{Goon:2012if}.
\end{itemize}
%

%\mbox{$\Delta$m$^2$ $\approx$ 1 eV$^2$}

%\cite{LBNO}
%A. Stahl et al. Expression of Interest for a very long baseline neutrino oscillation experiment (LBNO). CERN-%SPSC-2012-021, SPSC-EOI-007, CERN, 2012.

%\cite{nuSTORM}
%P. Kyberd et al. Letter of Intent: Neutrinos from STORed Muons. arXiv, June 2012.

%\cite{HyperK}
%K. Abe et al. Letter of Intent: The Hyper-Kamiokande Experiment Detector Design and Physics Potential. arXiv, %2011.

% \cite{ICARUS}
%A. Antonello et al. Search for ÓanomaliesÓ from neutrino and anti-neutrino oscillations at Æm2 Å 1 eV2 with muon %spectrometers and large LAr-TPC imaging detectors. CERN-SPSC-2012-010, SPSC-P-347, CERN, 2012.

% \cite{LBNE}
%J. Goon et al. Long-baseline neutrino experiment (lbne) project. Conceptual design report, October 2012.

\footnotetext {The Laguna-LBNO project is an FP7 project approved and funded until 2014. The main deliverable is a Technical Design Proposal which will outline the physics potential, technical feasibility, costing and siting of a long baseline neutrino oscillation experiment.}

\begin{figure}[hbt]
\centering
\includegraphics*[width=0.60\textwidth]{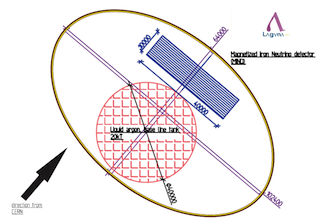}
\caption{The Laguna-LBNO far detectors Glacier and MIND in one of the caverns at the Pyh$\ddot{\mbox{a}}$salmi  complex.}
\label{lbno-fd}
\end{figure}

\begin{figure}[hbt]
\centering
\includegraphics*[width=0.60\textwidth]{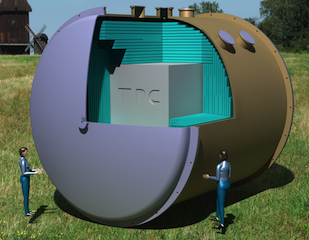}
\caption{Sketch of the Laguna-LBNO near detector consisting of a 20 bar Argon gas pressure vessel enclosing a time projection chamber (TPC) and surrounding plastic scintillator detector modules (TASD). Downstream of this pressure vessel, an electromagnetic calorimeter (ECAL) and magnetized iron neutrino detector (MIND) are planned for measurements of high energy events ($>3$ GeV).}
\label{lbno-nd}
\end{figure}

For the majority of the above proposals, plastic scintillators as trackers or tracking calorimeters with/without magnetic fields are an important part of the main detector options. For example at LBNO these plastic scintillators will instrument near and far detector components such as the magnetized iron detectors and the surroundings of the magnetized gas Argon TPC, Figures \ref{lbno-fd} and \ref{lbno-nd}. The use of solid polystyrene-based scintillators is widespread amongst the neutrino physics community, latest relevant examples include:
\begin{itemize}
\item the MINOS detectors (near and far),
\item the T2K near detectors  (off-axis ND280, on-axis INGRID),
\item the MINER$\nu$A detector.
\end{itemize}	 

%..............................................................................................................................
%..............................................................................................................................

%..............................................................................................................................
%..............................................................................................................................

%..............................................................................................................................
\begin{figure}[hbt]
\centering
\includegraphics*[width=0.60\textwidth]{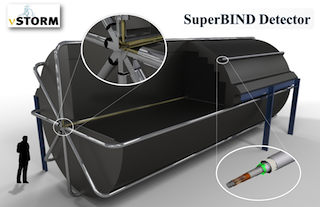}
\caption{The nuSTORM far detector SuperBIND.}
\label{nuSTORM-SuperBIND}
\end{figure}
%..............................................................................................................................

%..............................................................................................................................
\begin{figure}[hbt]
\centering
\includegraphics*[width=0.60\textwidth]{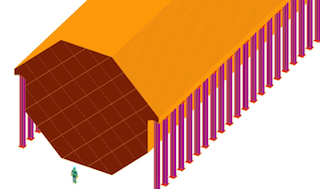}
\caption{The Neutrino Factory MIND.}
\label{NeutrinoFactory}
\end{figure}
%..............................................................................................................................

The large-scale detectors planned for future facilities, and requirements for ever increasing resolution and hence finer segmentation of the plastic scintillators lead to ever greater numbers of channels, with the usual need for cost/channel optimization.

Along with the performance of detectors obtained online, dedicated test runs have provided valuable information, for example the calibration detector tests or CalDet carried out for MINOS at the CERN-PS, which measured the energy resolution for the setup to be $21.42\%/$$\sqrt{E}$ (GeV), $56.6\%/$$\sqrt{E}$ and $56.1\%/$$\sqrt{E}$ (GeV) for electrons, protons and pions respectively. Much of this information is useful in designing future detectors although different readout schemes going from the MINOS detector photomultiplier tubes to silicon photomultipliers, and different geometries with finer segmentation call for further testing.
Some of the characteristics of the above detectors are poorly known, either because they have not been tested or because the conditions under which tests were carried out were not fully representative of the operational environment (e.g. no B-field). The evolution of software tools such as reconstruction algorithms also calls for a comparison to experimental data for validation. There is therefore a strong incentive to study:
\begin{itemize}
\item TASD: Stopping properties of pions and muons. Test beam: 200 MeV/c . This will be studied with the MICE EMR detector.
\item TASD: Electron and muon charge separation inside a magnetic field, in particular electron charge ID in electron neutrino interactions for the platinum channel at a neutrino factory. Test beam: 0.5 $\rightarrow$ 5 GeV/c. 
\item MIND: Muon charge identification, for wrong sign muon signature of a neutrino oscillation event: golden channel at a neutrino factory: requires correct sign background rejection of 1 in $10^4$, 0.5 $\rightarrow$ 5 GeV/c.
\item MIND: Hadronic shower reconstruction for identification of charged current neutrino interactions and rejection of neutral current neutrino interactions. Test beam: protons/pions 0.5 $\rightarrow$ 9 GeV/c.
\end{itemize}

\clearpage

\clearpage

\section{Motivation for tests}
The main motivation for test beam activities on prototype TASD and MIND detectors is to further understand how to optimize these detectors both in terms of performance and cost:
\begin{itemize}
\item{Performance: the intrinsic detector performance measured in terms of position, energy resolution must be coupled to our ability to extract relevant information such as momentum resolution, charge identification efficiencies etc... A better understanding of the interface between detector operation and reconstruction software is required, in particular at the level of the digitization process i.e. the description of what constitutes a hit in the reconstruction software. Several tools are already available and will benefit from experimental data. For example, the latest sterile neutrino analysis for nuSTORM is based on multi-variate analysis (MVA), which must be benchmarked against experimental data.}
\item{Cost: The costs of components that make up TASD and MIND detector modules evolve in time. Robust costing models are required to evaluate the cost of the large MIND-type detectors planned for nuSTORM or the Neutrino Factory. Although the cost of some of the passive components such as the steel is variable, and in particular the engineering costs for the steel, some larger variations can be observed for components such as photosensors where more manufacturers bring new products to the market.}
\end{itemize}

Extensive studies of MIND-type detectors have been carried out in the context of the Neutrino Factory. $\nu_\mu$ Charged Current interactions in such detectors leave a clear signature, in the form of a muon which has a clear track, penetrating several layers of steel, plus a far more limited in extent hadronic shower from hadronic interactions at the vertex.  This is the so-called "Golden Channel" at a Neutrino Factory. MIND detectors are self-calibrating detectors in that the magnetic field provides a measurement of momentum by bending, which can be checked against a momentum measurement by range for those muon events that are fully contained in the detector. Event reconstruction is relatively straightforward and consists in selecting tracks above a certain threshold in length, say 100 cm, and analyzing the vertex, extracting calorimetric information from the vertex to reconstruct the hadronic component of the interaction. A comprehensive study of this channel was carried out a decade ago for a Neutrino Factory configuration with high energy $\nu_\mu$, where 99.2\% of all $\nu_\mu$ CC events around 50 GeV produced tracks fulfilling the 100 cm length criteria \cite{Cervera:2000kp}. Further techniques were applied, using Multi-Variate Analysis to improve cuts on event selection \cite{Cervera:2010rz}. 

Another good example of the requirement for better knowledge of detector response comes from the search for sterile neutrinos. Data from some experiments measuring either oscillations from one neutrino type to another, or fewer counts than expected for a given neutrino species (disappearance) are not fully consistent with the Standard Model and its three lepton flavours. Other models have been proposed with a fourth (or more) species of neutrinos, the sterile neutrinos. These models describe the oscillation of neutrinos from one flavor to another via sterile neutrinos, or oscillations into sterile states. A number of experiments have also reported data which do not support the existence of sterile neutrinos. 

\begin{figure}[hbt]
\centering
\includegraphics*[width=0.50\textwidth]{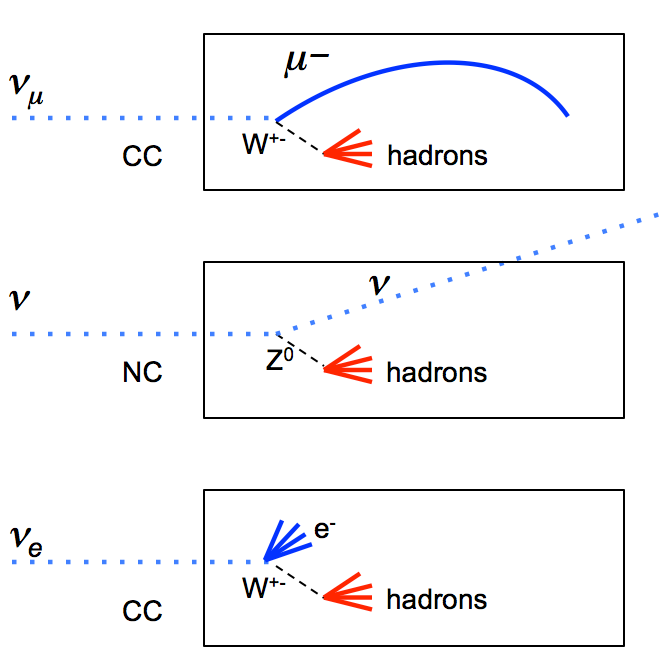}
\caption{Sketch of event topologies in the MIND. The dashed lines with the W/Z bosons are for illustration only and do not represent tracks.}
\label{event_topo}
\end{figure}

Several proposals are under consideration worldwide to address the sterile neutrino issue. One such proposal is nuSTORM which plans to produce pure beams of neutrinos of a given flavor, from the two stage decay $\pi^+$ $\rightarrow$ $\mu^+$ + $\nu_\mu$ and $\mu^+$ $\rightarrow$ $e^+$ + $\nu_e$ + $\overline{\nu}_\mu$ or the corresponding (anti)leptons from a $\pi^-$ beam. Typical event topologies for different neutrino interactions in a MIND-type detector are shown in Figure \ref{event_topo}. Events of interest in a MIND are charged current (CC) interactions resulting in a lepton in the final state. As can be seen, $\nu$$_e$ events are far more challenging to reconstruct in the MIND, due to the much shorter tracks and similarity with hadronic energy deposition. At nuSTORM, events with a clear muon track are those that are ultimately retained for analysis. For a $\mu^+$ beam, although the appearance channel  $\nu$$_e$$\rightarrow$$\nu$$_\mu$ is doubly suppressed in comparison with the disappearance channel, the backgrounds are more manageable. The appearance channel $\nu$$_e$$\rightarrow$$\nu$$_\mu$ is the "golden" channel in the search for sterile neutrinos at nuSTORM.

%..............................................................................................................................
%..............................................................................................................................

\section{Baby MIND: The Magnetised Iron Neutrino Detector prototype}
The baby MIND detector to be tested is a 50 tonne prototype for a full MIND-type detector as is planned for the LBNO detector (downstream of both the gas argon TPC at the near detector and the LAr LEM-TPC at the far detector) and $\nu$STORM near and far detectors. Dimensions are shown in Figures \ref{mindsketch} \& \ref{mindplanes}, MIND parameters are listed in Appendix B and are subject to further optimisation. Assuming a minimum ionising muon looses 11.4 MeV/c per cm of steel, 51 plates of 3 cm-thick steel interleaved with 1.5 cm-thick modules of plastic scintillator would contain 2 GeV/c muons.
Due to their cost, the amount of iron is limited by mass to ~50 t, and the number of plastic scintillator modules is limited to 50. The characteristic radiation length in iron is $X_0$ = 13.8 g.cm$^{-2}$ corresponding to 1.75 cm, or 87 radiation lengths for an iron depth of 153 cm. The characteristic nuclear interaction length in iron is $\lambda$$_I$ = 132.1 g.cm$^{-2}$ corresponding to 16.78 cm, or 9 interaction lengths for an iron depth of 153 cm. Although the energy resolution degrades with increasing plate thickness in the range 1-5 cm, the number of sampling points (scintillator modules) would have to be larger than the limit of 50 for this detector, especially if going for the same iron depth of 153 cm. As an example, for a 1 cm iron plate thickness, with 50 modules, the total iron depth is 51 cm and the radiation and interaction lengths scale accordingly. For larger plate thicknesses, the number of sampling points decreases (31 sampling points for 5 cm iron plates). Given the limits on iron and number of scintillator modules, the iron thickness was chosen to be 3 cm. This value is also retained for the neutrino factory studies.

%..............................................................................................................................
\begin{figure}[hbt]
\centering
\includegraphics*[width=0.7\textwidth]{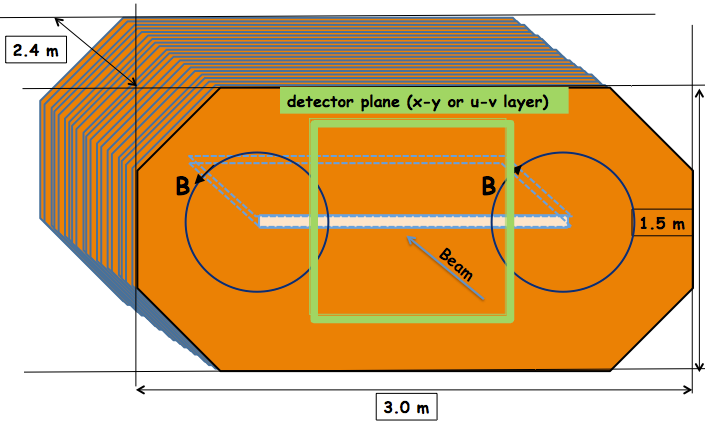}
\caption{Sketch of the Magnetised Iron Neutrino Detector (baby MIND) prototype planned for tests at the SPS.}
\label{mindsketch}
\end{figure}
%..............................................................................................................................

%..............................................................................................................................
\begin{figure}[hbt]
\centering
\includegraphics*[width=0.5\textwidth]{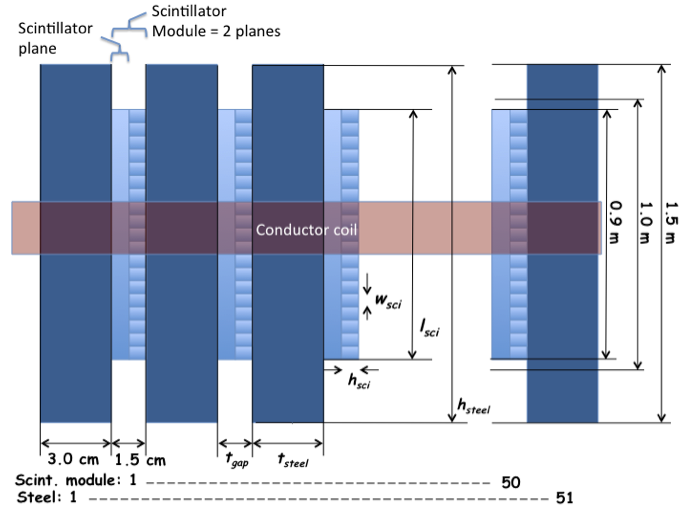}
\caption{Sketch showing dimensions of the iron slabs and plastic scintillator modules for the baby MIND prototype.}
\label{mindplanes}
\end{figure}
%..............................................................................................................................

\clearpage

\subsection{MIND prototype simulations}
Preliminary simulations in Geant4 of a MIND prototype are reported here. Further optimisation is ongoing, particularly concerning the geometry, definition of the magnetic field and reconstruction algorithms for pions. Assumptions taken for the geometry are octagonal plates 2 m $\times$ 1 m, two transmission lines, with a 2.8 m detector depth (along the beam axis). 
Four different scenarios were tested to validate the steel thickness and for an indication of whether the choice of scintillator geometry is acceptable (0.7 cm high rectangular bars vs 1.7 cm high triangular bars), scintillator pitch = 1.0 cm in all scenarios:
\begin{itemize}
\item 3 cm steel plate, 1.5 cm scintillator module (i.e. 0.75 cm X plane + 0.75 cm Y plane),
\item 2 cm steel plate, 1.5 cm scintillator module,
\item 3 cm steel plate, 3.5 cm scintillator module,
\item 2 cm steel plate, 3.5 cm scintillator module.
\end{itemize}

In assessing the various efficiencies from simulations for a small MIND prototype exposed to a charged particle beam, the following were taken into consideration:
\begin{itemize}
\item{a) The total number of tracks (or simulated particles in the detector);}
\item{b) The number of tracks reconstructed (using the Kalman filter);}
\item{c) The number of successful tracks (where successful means that the correct charge is identified).}
\end{itemize}

The efficiencies are then defined as:
\begin{itemize}
\item{1) The reconstruction efficiency is b/a.}
\item{2) The charge identification efficiency is c/a.}
\end{itemize}

\subsubsection{Muon reconstruction efficiencies}
\begin{figure}[hbt]
\centering
\includegraphics*[width=0.24\textwidth]{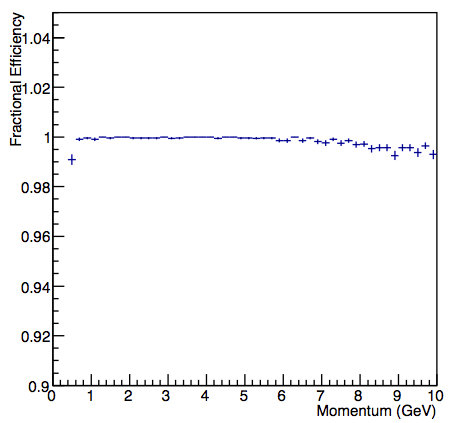}
\hfill
\includegraphics*[width=0.24\textwidth]{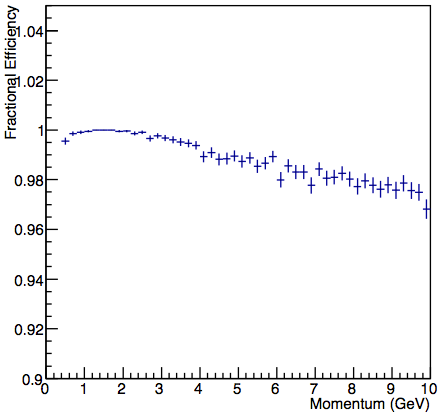}
\hfill
\includegraphics*[width=0.24\textwidth]{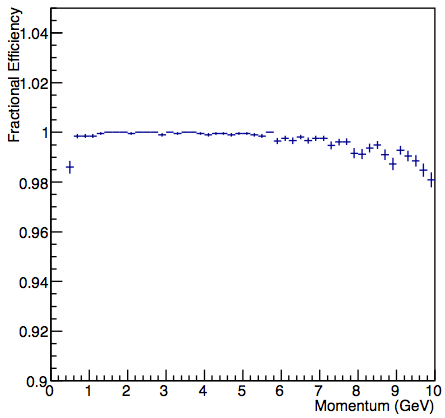}
\hfill
\includegraphics*[width=0.24\textwidth]{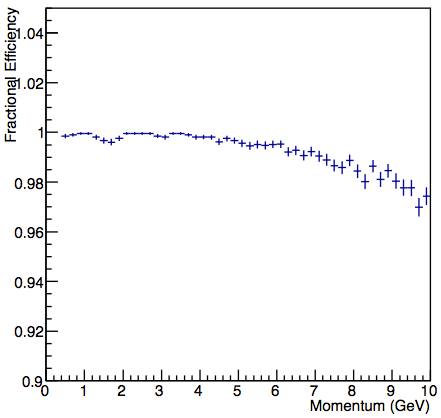}
\caption{Reconstruction efficiencies for $\mu$$^+$ for different steel and scintillator combinations, clockwise from top left: a) 3 cm steel, 1.5 cm scintillator. b) 2 cm steel, 1.5 cm scintillator. c) 3 cm steel, 3.5 cm scintillator. d) 2 cm steel, 3.5 cm scintillator.}
\label{mindmuonefficiency}
\end{figure}

Muon reconstruction efficiencies are shown in Figure \ref{mindmuonefficiency}. All four combinations of steel and scintillator thicknesses show good efficiencies at low momenta $<$2 GeV/c. The combination showing the best performance over the widest momentum range is 3.0 cm of steel and 1.5 cm of scintillator, with efficiencies close to 100\% up to 6 GeV/c, staying above 99\% up to 10 GeV/c. These efficiencies remain good for the other scenarios, dropping to 97\% at high momenta.
Charge identification efficiencies are identical for all scenarios, close to 100 \% for all momenta above 1 GeV/c.
%..............................................................................................................................
%..............................................................................................................................

A small fraction of events are not successfully reconstructed for the baby MIND prototype. Track reconstruction using the Kalman filter from RecPack is based on a number of criteria, the most important being the number of hits along the track and the apparent curvature of the track (i.e. is the curvature well enough defined as to determine the momentum). The Kalman filter algorithm allows for the propagation of track parameters back through successive detector planes using a helix model that considers multiple scattering and energy loss. In a neutrino detector where a $\nu_\mu$ charged current interaction leads to hadronic activity at the interaction vertex in addition to an outgoing muon, the filter works well when the muon travels much further than particles related to the hadronic interactions. By ranking hits as a function of distance from the interaction vertex, it is possible to distinguish hits furthest away from this vertex as due only to a muon. Those hits then act as a seed for the Kalman filter. A typical reconstruction analysis will consider a number of planes (e.g. five) furthest downstream where hits are due to muons only. At high momenta, the technique works well. It has however limitations at high $Q^2$ or low neutrino energy, when the muon range is comparable to the range of the hadronic activity and identification of the muon becomes more challenging. When the Kalman filter fails, a cellular automaton method can be applied. It forms possible trajectories from a ranking of hits using a neighbourhood function. At very low muon momenta, $E_\mu < 0.5$ GeV, track reconstruction is limited when angular deflections due to multiple scattering dominate over those due to the magnetic field in the steel.

In a charged particle beam scenario with a reduced detector depth along the charged particle beam axis, the Kalman filter can fail to reconstruct a track at low momentum if the number of hits is insufficient, or at high momentum if the curvature of the track is insufficient. Analyses for the much larger Neutrino Factory MIND or nuSTORM SuperBIND with true neutrino interactions yield efficiencies which are much closer to 100\%, with a few events ($<<1\%$) failing reconstruction due to significant scattering affecting track curvature. For these detectors, pattern recognition and event reconstruction are more complex, and merit some comparison with test beam data.

\begin{figure}[hbt]
\centering
\includegraphics*[width=0.24\textwidth]{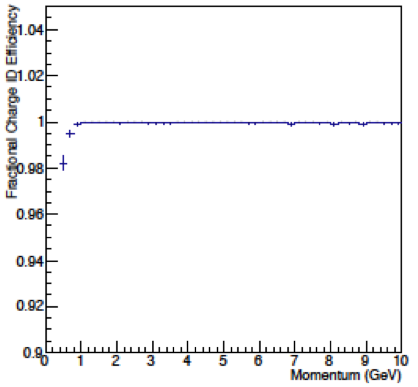}
\hfill
\includegraphics*[width=0.24\textwidth]{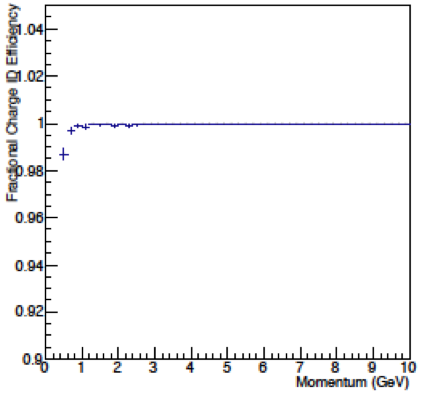}
\hfill
\includegraphics*[width=0.24\textwidth]{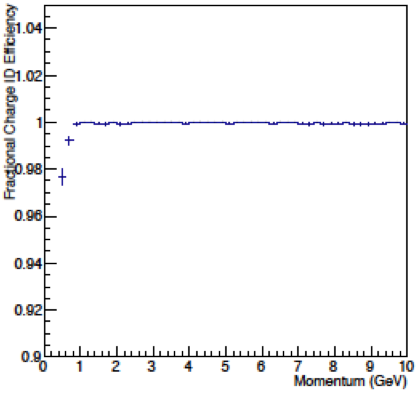}
\hfill
\includegraphics*[width=0.24\textwidth]{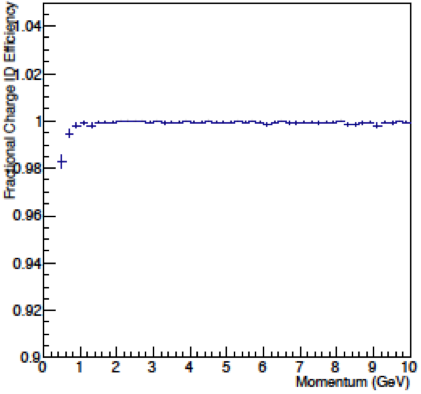}
\caption{Charge identification efficiencies for $\mu$$^+$ for different steel and scintillator combinations, clockwise from top left: a) 3 cm steel, 1.5 cm scintillator. b) 2 cm steel, 1.5 cm scintillator. c) 3 cm steel, 3.5 cm scintillator. d) 2 cm steel, 3.5 cm scintillator.}
\label{mindmuchargeID}
\end{figure}

Charge identification efficiencies are similar for all scenarios, close to 100 \% for all momenta above 1 GeV/c, Figure \ref{mindmuchargeID}. Detailed simulations were not carried out below 1 GeV/c. It is to be expected that the thickness of steel and scintillator start playing a role because of multiple scattering. A hint of the fall in charge identification efficiencies can be seen in Figure \ref{mindmuchargeID} with a couple of data points below 1 GeV/c where the thinnest steel and plastic scintillator combination has marginally better efficiency.
Pion track reconstruction is more challenging because of the hadronic shower development. The pion reconstructed momenta show that individual tracks are poorly reconstructed, Figure \ref{mindpionreconeff}. The input momentum distribution was uniform in pz between 0.3 GeV/c and 10 GeV/c. The reconstructed momentum distribution is peaked at low momenta. Comparison of $\pi$$^+$ and $\pi$$^-$ shows some discrimination of charge although further analysis is required.
Charge identification efficiencies are poor for pions, Figure \ref{mindpionchargeID}. Single track charge identification is unreliable and requires further work.

%..............................................................................................................................
%..............................................................................................................................

\subsubsection{Pion reconstruction efficiencies}

\begin{figure}[hbt]
\centering
\includegraphics*[width=0.24\textwidth]{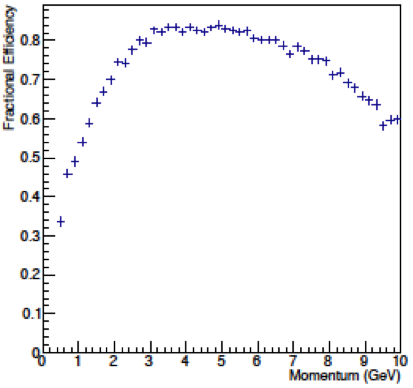}
\hfill
\includegraphics*[width=0.24\textwidth]{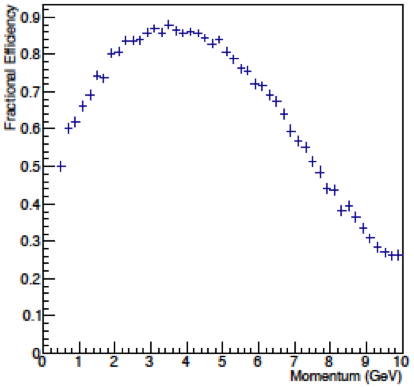}
\hfill
\includegraphics*[width=0.24\textwidth]{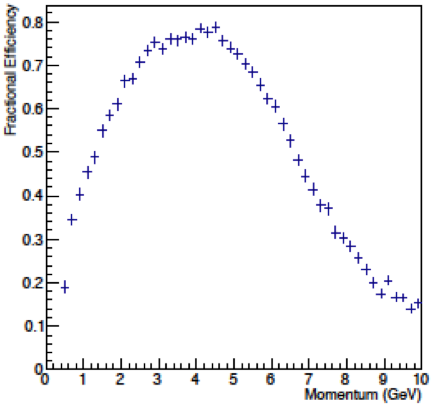}
\hfill
\includegraphics*[width=0.24\textwidth]{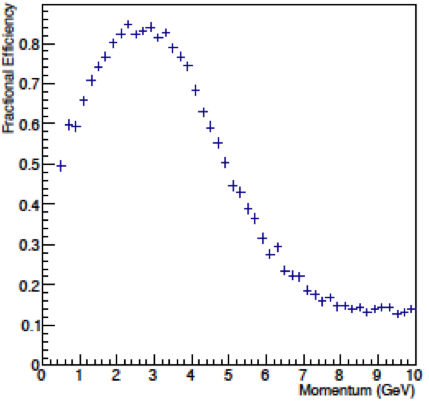}
\caption{Reconstruction efficiencies for $\pi$$^+$, clockwise from top left: a) 3 cm steel, 1.5 cm scintillator. b) 2 cm steel, 1.5 cm scintillator. c) 3 cm steel, 3.5 cm scintillator. d) 2 cm steel, 3.5 cm scintillator.}
\label{mindpionreconeff}
\end{figure}

\begin{figure}[hbt]
\centering
\includegraphics*[width=0.24\textwidth]{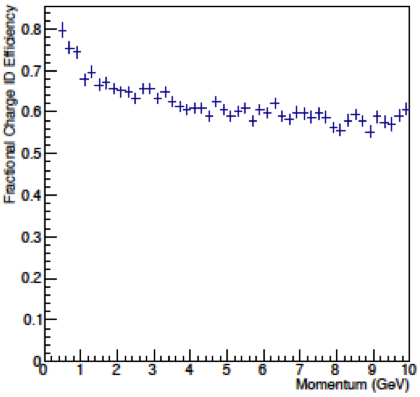}
\hfill
\includegraphics*[width=0.24\textwidth]{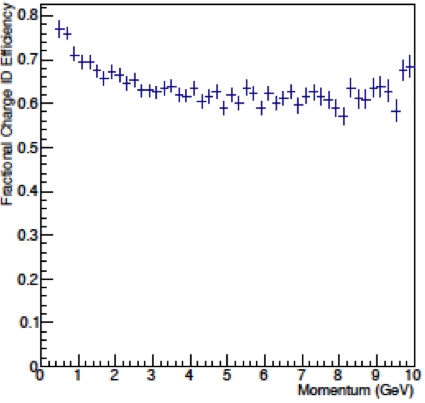}
\hfill
\includegraphics*[width=0.24\textwidth]{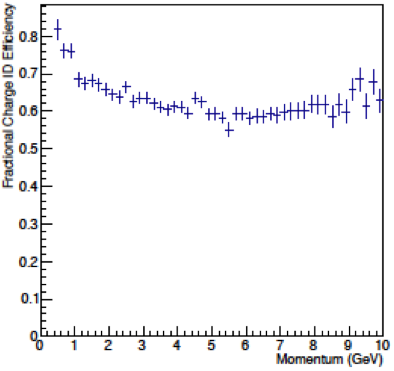}
\hfill
\includegraphics*[width=0.24\textwidth]{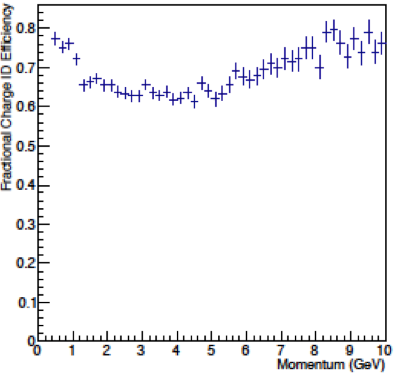}
\caption{Charge identification efficiencies for $\pi$$^+$, clockwise from top left: a) 3 cm steel, 1.5 cm scintillator. b) 2 cm steel, 1.5 cm scintillator. c) 3 cm steel, 3.5 cm scintillator. d) 2 cm steel, 3.5 cm scintillator.}
\label{mindpionchargeID}
\end{figure}

Pion track reconstruction is more challenging because of the hadronic shower development. The pion reconstructed momenta show that individual tracks are poorly reconstructed, Figure \ref{mindpionreconeff}. The input momentum distribution was uniform in $p_z$ between 0.3 GeV/c and 10 GeV/c. The reconstructed momentum distribution is peaked at low momenta. Comparison of $\pi$$^+$ and $\pi$$^-$ shows some discrimination of charge although further analysis is required.
Charge identification efficiencies are poor for pions, Figure \ref{mindpionchargeID}. Single track charge identification is unreliable and requires further work.

%..............................................................................................................................
%..............................................................................................................................

%..............................................................................................................................
%..............................................................................................................................

\subsection{MIND prototype magnetisation}
Concerning the magnetization of the MIND prototype, a low carbon steel will be selected. There are no particular radiation or environmental constraints (corrosion/humidity). The magnetisation will be set by passing a current through one or more conductor coils. Specifications for the field are the following:
\begin{itemize}
\item{field value: 1.5 T $\pm20\%$,}
\item{knowledge of field in volume of interest to a precision of 1e-4, especially Bx and By components,}
\item{field uniformity within steel along projection of plastic scintillator volume: 10\%,}
\item{field value outside MIND volume: maximum = 10 mT.}
\end{itemize}
The assumption made concerning power supplies is that one can be provided by CERN. The coil design will therefore be made as a function of available power supplies.
Initial studies were carried out to optimise the uniformity of the $B_y$ component of the field whilst minimising the $B_x$ component. Although this does not represent the current consensus on MIND-type detector design, this approach was meant to minimise uncertainties in the knowledge of the B-field. The resulting geometry led to a considerable height increase for the steel plates (factor $\times$2), in order to have a return path for the field lines well away from the detector plane area. The results of the optimisation of field lines is an impressive constraining of the $B_x$ component of the field, 115 Gauss (0.7\% of $B_y$) for the two coil configuration, compared to 8210 Gauss (110\% of $B_y$) for the one coil configuration. However, the doubling of the cost of steel and the introduction of large "empty" slots are disadvantages which drive the adoption of the more affordable and adequate one coil configuration.

\begin{figure}[hbt]
\centering
\includegraphics*[width=0.7\textwidth]{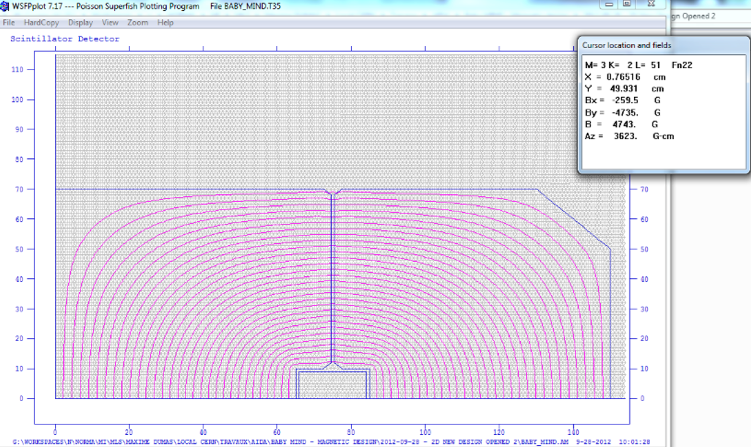}
\caption{Simulation of the magnetic field lines for the MIND prototype with a vertical gap for the insertion of a measuring device. This gap can be seen at ~ 75 cm on the horizontal axis. One possible method of measuring the B field is to insert a probe embedded in a non-magnetisable material into this vertical gap. Field lines in the detector module area between 0 and +45cm on both vertical and horizontal axes can be inferred from measurements made with the probe along the gap length.}
\label{mindfieldgap}
\end{figure}

Having chosen the one-coil configuration, attention is now turning to the challenge of determining with accuracy the value of the field, not simply the total field but separate knowledge of the $B_x$ and $B_y$ components of the field. One solution that has been proposed and which is currently being investigated is to create a slot away from the detector planes from the coil to the outer edge of the steel plate, into which a non-magnetised material such as a stainless steel or aluminium is inserted with an embedded magnetic field sensor. By displacing this sensor along the entire length of the gap, it is possible to create a map of the field in the gap and thus infer the field lines in the area of the detector planes, Figure \ref{mindfieldgap}. This technique could have relevance for the much larger MIND-type detectors planned for future facilities.
Further optimisation work is required to produce a design for the MIND that includes detailed maps of the B- field, thorough mechanical design and integration of a B-field measurement system.

%..............................................................................................................................
%..............................................................................................................................

\subsection{MIND prototype tests}
Online testing will address the full characterisation of the MIND prototype detector, assessing its energy and spatial resolution. Comparisons will be drawn with prior simulation work. One of the more significant elements to be checked in the simulations will be the digitisation, which requires hardware efficiencies that are measured online.

\clearpage

\section{The Totally Active Scintillator Detector (TASD) prototype}

The TASD detector consists of 50 modules of plastic scintillators. Each module is instrumented with one X and one Y plane, with 90 scintillator bars per plane. The bar width, height and length are 1.0 cm, 0.7 cm and 90 cm respectively. The distance between modules can be varied from 0 to 2.5 cm. Other components such as active detectors or passive sheets of material can be inserted in these 2.5 cm gaps if required. The full detector depth can therefore be varied from 75 cm to ~200 cm and in its compact form, it is ~1 $m^3$ in volume. Geometric data are presented in Appendix A.
Separation by range of pions and muons of same momenta is possible at low momenta. The range in plastic of 280 MeV/c pions and muons is ~60 cm and ~72 cm respectively. 

The goal is to have a software framework with a set of simulation tools that can:
\begin{itemize}
\item describe all the physics from beam parameters to interactions within the detector.
\item accurately reproduce the performance of the detectors: especially optical photon transport. 
\item enable an extrapolation from test beam conditions to real detector scenarios.
\end{itemize}
\subsection{TASD on the MICE beamline at RAL [UK]}

\begin{figure}[hbt]
\centering
\includegraphics*[width=0.32\textwidth,height=100pt]{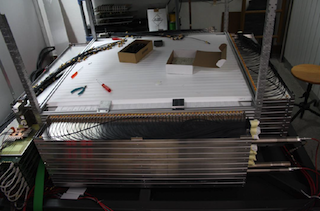}
\hfill
\includegraphics*[width=0.32\textwidth,height=100pt]{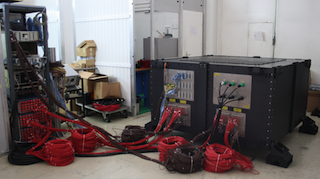}
\hfill
\includegraphics*[width=0.32\textwidth,height=100pt]{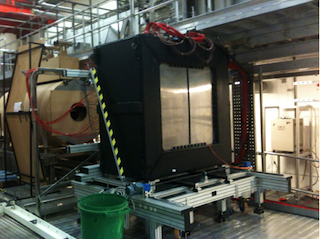}
\caption{Commissioning of the MICE Electron Muon Ranger (EMR) in 2013: left) Assembly of the MICE-EMR detector at the University of Geneva, photo taken in June 2013, middle) Completed MICE-EMR, photo taken mid-September 2013, right) The MICE-EMR installed at the end of the MICE beamline at the Rutherford Appleton Laboratory (RAL) in the UK, photo taken 27th September 2013.}
\label{MICE-EMR}
\end{figure}

\begin{figure}[hbt]
\centering
\includegraphics*[width=0.5\textwidth]{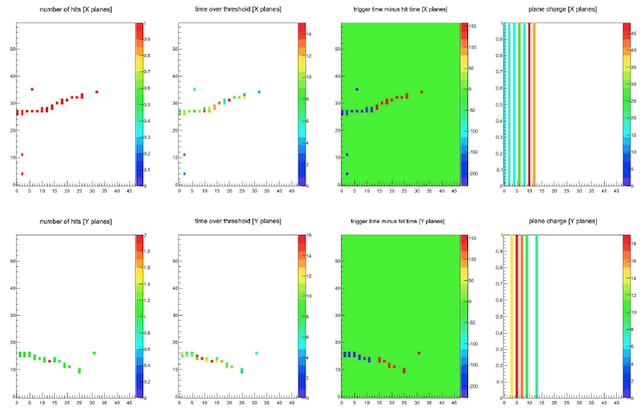}
\caption{MICE-EMR event display showing a 140 MeV/c negative muon entering the detector from the left, recorded during test beams on the MICE beamline in October 2013.}
\label{EMR-event}
\end{figure}

A Totally Active Scintillator Detector (TASD) was designed and constructed over several years at the University of Geneva. It was commissioned in summer 2013, and shipped to the Rutherford Appleton Laboratory (RAL) in the UK where it was installed at the end of the Muon Ionization Cooling Experiment (MICE) beamline on 27th September 2013, Figure \ref{MICE-EMR}. Considerable experience has been gained at the University of Geneva with all stages of the realization of such detectors.
First online test beam runs were carried out in October 2013, showing excellent particle identification capabilities of the TASD for low momentum ($< 400$ MeV/c) muons, pions and electrons. Data analysis is ongoing, an event display is shown in Figure \ref{EMR-event}

%..............................................................................................................................
%..............................................................................................................................
%..............................................................................................................................
%..............................................................................................................................

\clearpage

\subsection{TASD simulations}
\begin{figure}[hbt]
\centering
\includegraphics*[width=0.35\textwidth]{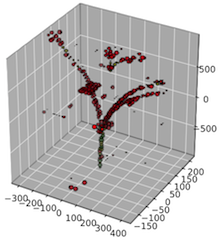}
\hfill
\includegraphics*[width=0.35\textwidth]{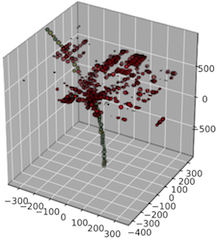}
\caption{3D hit maps in the TASD with basic digitisation described previously for 5 GeV electrons left), and protons right).}
\label{tasd3D}
\end{figure}

Preliminary work has been carried out to exercise some of the simulation tools for the TASD, using a basic version of the event generator, Geant4 for particle transport, and basic digitisation. Track reconstruction is work in progress, planned for a later stage in the project.
Elements of the detector geometry are parameters directly implemented in Geant4. The digitisation works by summing the energy deposited in each scintillator bar, with a poisson distribution around a mean of 15 photo-electrons/(1.8 MeV). It is planned to include light collection attenuation, and a more detailed description of SiPM response in the digitisation.
Hit maps for 5 GeV electrons, muons and protons are shown in Figure \ref{tasd3D}. The TASD detector here is in a configuration with a 2.5 cm gap between the plastic scintillator modules. The work required for track reconstruction can be appreciated. Tracking of the primary electron and secondaries produced via interactions in the TASD will require additional work. Simulations at lower momenta are also foreseen, down to 0.5 GeV/c, especially relevant for electron and muon charge identification.

%..............................................................................................................................
%..............................................................................................................................

Much of the offline R\&D will be carried out on the components of the TASD such as the plastic scintillators, wavelength shifting fibers, photosensors and associated electronics. Since these components are shared with the MIND detector prototype, they are described in a joint section further in this document. 

The main requirement concerning the beam is that the TASD be placed inside a large aperture magnet. The TASD dimensions were decided based on the assumption that it would be located in the Morpurgo magnet on the H8 beam line in the north area. Further details are provided in the "beam request" section.

\clearpage

\section{MIND and TASD shared detector modules}
Having a MIND prototype downstream of the TASD and operating both detectors online at the same time had been envisaged but the high cost of duplicating the detector modules was prohibitive. Therefore, the same plastic scintillator detector modules will be deployed in the TASD and MIND detector prototypes, which means that the two detectors cannot be operated at the same time. The MIND part of the near detector can be used as a prototype of the far detector and installed in the North area extension behind the liquid Argon prototype.

\subsection{Scintillators}
\begin{figure}[hbt]
\centering
\includegraphics*[width=0.4\textwidth]{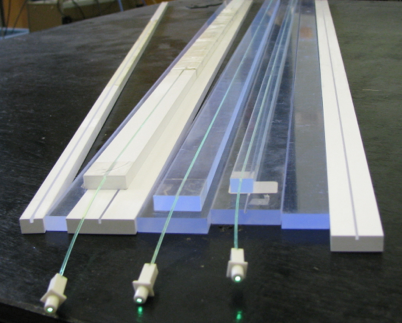}
\hfill
\includegraphics*[width=0.5\textwidth]{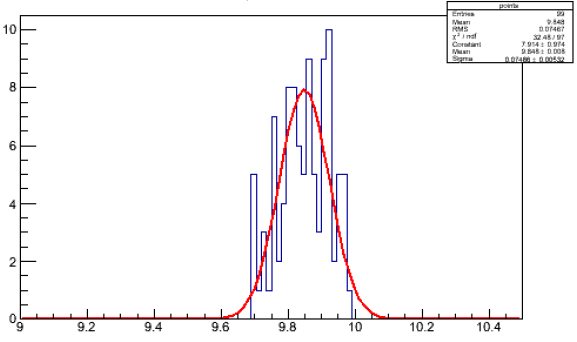}
\caption{Left: Prototype plastic scintillators of different dimensions produced by Uniplast. Right: Distribution of measured width for 100 randomly selected plastic scintillator bars for the chosen width of 10 mm, the X scale is [mm].}
\label{scintillator}
\end{figure}

The plastic scintillator bars will be supplied by the Institute for Nuclear Research (INR) of the Russian Academy of Sciences. The nominal parameters for the geometry are bars of 90 cm long, 0.7 cm in height and 1.0 cm in width, examples are shown in Figure \ref{scintillator}. A small batch of prototypes has been manufactured by Uniplast based in Vladimir (Russia) and shipped to Geneva for testing. These extruded scintillator slabs are polysterene-based with 1.5\% of paraterphenyl (PTP) and 0.01\% of POPOP, similar to the plastics used for the T2K SMRD detector counters. The surface is etched with a chemical agent (Uniplast) to create a 30-100 $\mu$m layer acting as a diffusive reflector. Slabs of three different sizes have been manufactured (895 $\times$ 7 $\times$ 10 mm$^3$, 895 $\times$ 7 $\times$ 20 mm$^3$, 895 $\times$ 7 $\times$ 30 mm$^3$) with 2 mm deep grooves  of different widths (1.1 mm, 1.3 mm or 1.7 mm) to embed optical fibres of different diameters.

%..............................................................................................................................
%..............................................................................................................................

\begin{figure}[hbt]
\centering
\includegraphics*[width=0.4\textwidth]{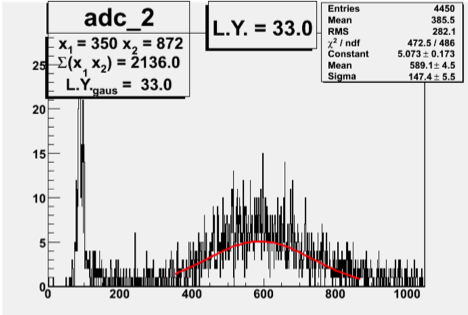}
\hfill
\includegraphics*[width=0.4\textwidth]{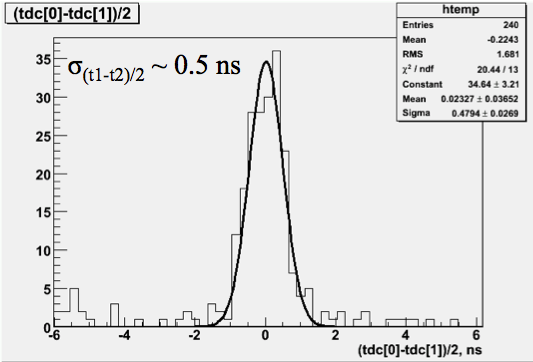}
\caption{Response of a scintillator slab and silicon photomultiplier to a minimum ionising particle: a) light yield and b) timing properties.}
\label{scintillatormip}
\end{figure}

\begin{figure}[hbt]
\centering
\includegraphics*[width=0.35\textwidth]{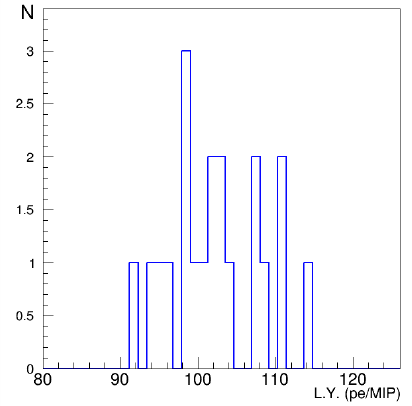}
\includegraphics*[width=0.45\textwidth]{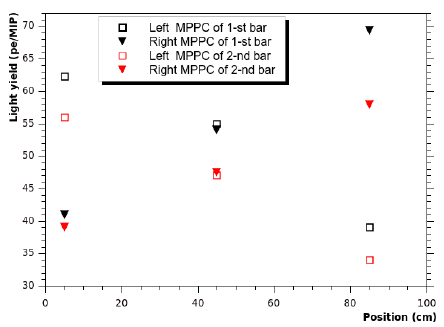}
\caption{Light yield from a 0.7 cm-thick plastic scintillator bar, readout from both ends. The plot on the left shows the sum of the two photosensors. The plot on the right shows the individual contribution of each of the two photosensors to the light yield as a function of position along the bar. The photosensors used for these tests were of the T2K-type, Hamamatsu MPPCs 1.3 $\times$ 1.3 mm$^2$ sensitive area.}
\label{lightyield}
\end{figure}

Tests were carried out at INR to determine basic light yield and timing properties. A wavelength shifting fiber (WLS) from Kuraray (200 ppm, S-type) of d = 1.0 mm was embedded into the 1.1 mm wide groove with a silicon grease (TSF451-50M) to improve optical contact between the scintillator groove surface and the fiber. Hamamatsu MPPC photosensors (1.3 $\times$ 1.3 mm$^2$, 667 pixels, 50 $\times$ 50 $\mu$m$^2$,  gain = 7.5 $\times$ 10$^5$ @$25\,^{\circ}\mathrm{C}$) were connected to both ends of the ~1m long WLS fibers. A cosmic telescope was set up with two trigger counters. Measurements were made at the center of the scintillator slabs. The temperature during testing was 25-$28\,^{\circ}\mathrm{C}$.
Results are summarised in Table \ref{lightyields}. Typical response to a minimum ionising particle is shown in Figure \ref{scintillatormip}. Results show good light yield for all bar thicknesses, the highest light yield was obtained with the narrowest 10 mm width. Comparisons with/without chemical reflector show an increase of light yield of a factor 2.5 when the chemical reflector is present. The effect of the silicon grease is close to 60\%. For the final assembly, the silicon grease would be replaced by glue, which is expected to have roughly the same effect. An additional Tyvek reflector provides a 20\% increase in light yield, though this reflector is not currently planned for the prototype detectors. The light yield was measured to be 50 p.e./MIP on average for one photosensor, Figure \ref{lightyield}.
Timing properties were studied for the two-sided readout, combining both ends with the result: $\sigma$($(t_1-t_2)$/2) = 0.5 ns. The timing is mostly determined by the fiber decay constant, $\tau$$_{fiber}$ $\backsim$ 12 ns.

%..............................................................................................................................
%..............................................................................................................................

%..............................................................................................................................
%..............................................................................................................................  

%============================================
\begin{table}[h]
\centering
\caption{\em Light yields from cosmic tests with prototype scintillator bars of different widths, unit is the photoelectron.}
\begin{tabular}{cccccccc}
\toprule
\textbf{Bar width [mm]} &	\textbf{MPPC 1 [p.e.]}  & 	\textbf{MPPC 2 [p.e.]}  &	\textbf{Sum [p.e.]}\\
\hline
\multicolumn{4}{l}{Bar with no chemical reflector}\\
\hline
10	&	15.7	 &	15.8	&	31.5 \\		
20	&	15.5	 &	13.6	&	29.1 \\	
30	&	12.8	 &	11.5	&	24.3 \\
20 + Tyvek reflector 100-200 $\mu$m		&	41.8	 &	34.8	&	76.6 \\
\hline
\multicolumn{4}{l}{Bar with chemical reflector}\\
\hline		
10	&	46.0	 &	36.8	&	82.8 \\	
20 (1) w/o grease	&	25.7	 &	22.1	&	47.8 \\	
20 (1)	&	39.7	 &	35.7	&	75.4 \\	
20 (1) + Tyvek reflector	&	49.3	 &	44.0	&	93.3 \\	
20 (2)	&	32.6	 &	28.2	&	60.8 \\		
30	&	31.2	 &	26.6	&	57.8 \\	
\bottomrule
\end{tabular}
\label{lightyields}
\end{table}
%============================================

\clearpage

\subsection{Choice of photosensor}

The first silicon photomultiplier device used on a large scale in a physics experiment is the Hamamatsu MPPC S10362-13-050C instrumenting all scintillator detectors at the ND280 near detector complex of the T2K experiment \cite{Amaudruz:2012esa}. It was derived from a commercial device, with the sensitive area increased from 1 $\times$ 1 mm${^2}$ to 1.3 $\times$ 1.3 mm${^2}$ to provide better acceptance for the light emitted from the 1.0 mm diameter wavelength shifting fiber from Kuraray. This MPPC consists of 667 pixels, each working in limited Geiger mode with an applied voltage slightly above the breakdown voltage ($V_{bd}=70V$). With the production of a photo-electron in a pixel, a Geiger avalanche is generated, which is then passively quenched by a built-in resistor in each pixel. The induced charge is independent of the number of photo-electrons produced, and is directly proportional to the over voltage which is therefore a crucial parameter: $Q=C(V-V_{bd})$ or $Q=C{\delta}V$. In order to operate in the linear regime, where the MPPC output charge is directly proportional to the incoming photons, it is crucial that the total number of photo-electrons remains below the number of pixels in the device.

The thorough work done in the selection and characterization of photosensors for the T2K experiment serves as a very good basis for the selection of photosensors for the prototypes. Photosensor design is a fast evolving field. Several manufacturers offer a variety of products.  Table \ref{photosensorcomparison} lists characteristics and measured performance for a selection of devices from different manufacturers tested at the INR in Russia in 2013.

The first set of devices from the latest generation of photosensors from Hamamatsu were commercialized in 2013. Specifications from this manufacturer indicate improved dark noise, lower after pulse and higher PDE. Smaller cell sizes are also available, increasing the dynamic range for our application. A short comparative study was carried out, measuring the older generation photosensors from Hamamatsu used extensively at the T2K ND280 near detector against the newer generation photosensors. In a geometrical configuration with a photosensor active area of 1.3 $\times$ 1.3 mm$^2$ and 50 $\mu$m cell size, identical to the older generation MPPC, when summing the light from both devices either end of the bar, a light yield of 141 p.e. is obtained with the latest generation S12825-050C, compared with 100 p.e. for the older generation S10362-13-050C. The higher light yield is due to a higher photon detection efficiency (PDE).

\begin{figure}[hbt]
\centering
\includegraphics*[width=0.45\textwidth]{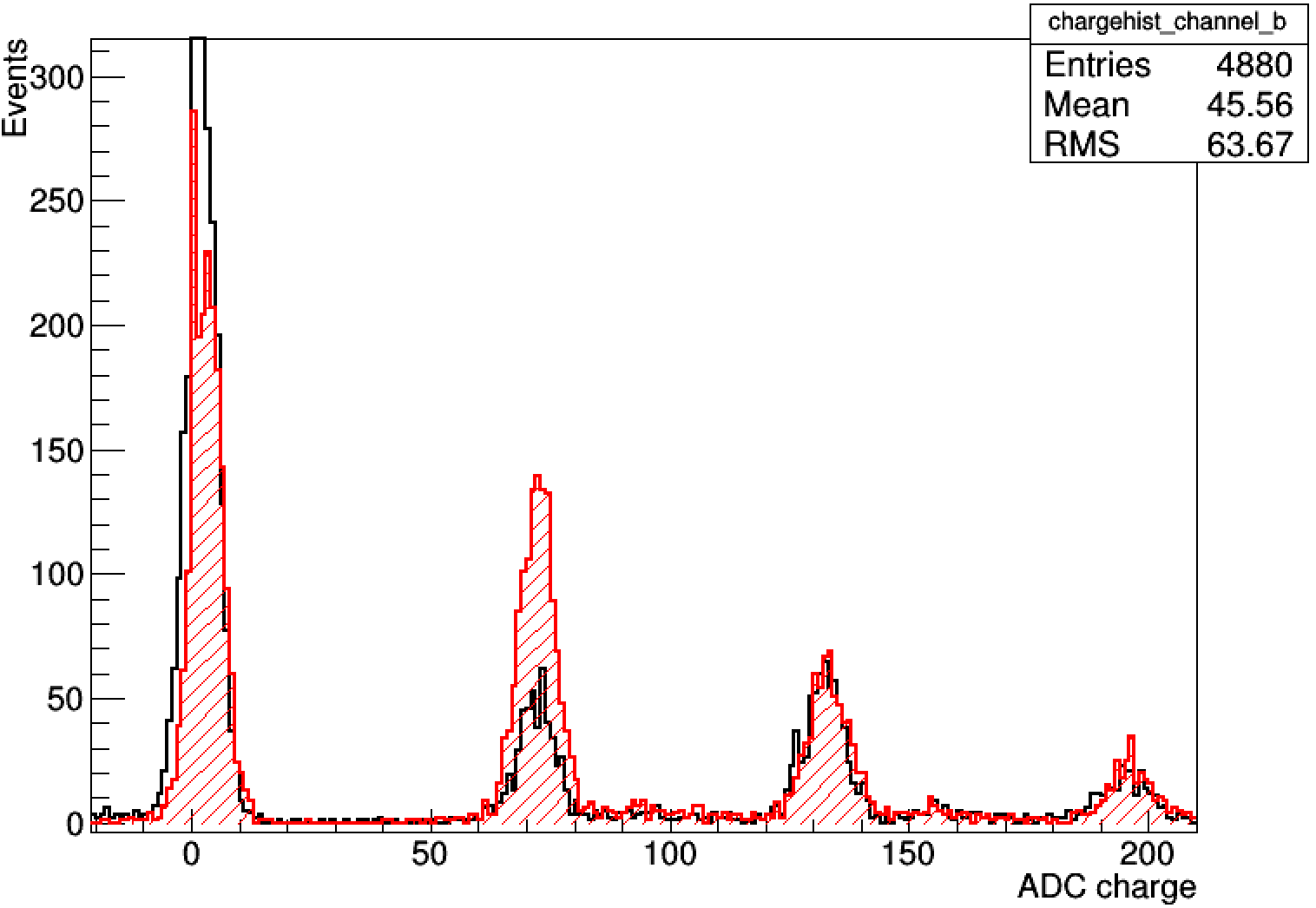}
\includegraphics*[width=0.45\textwidth]{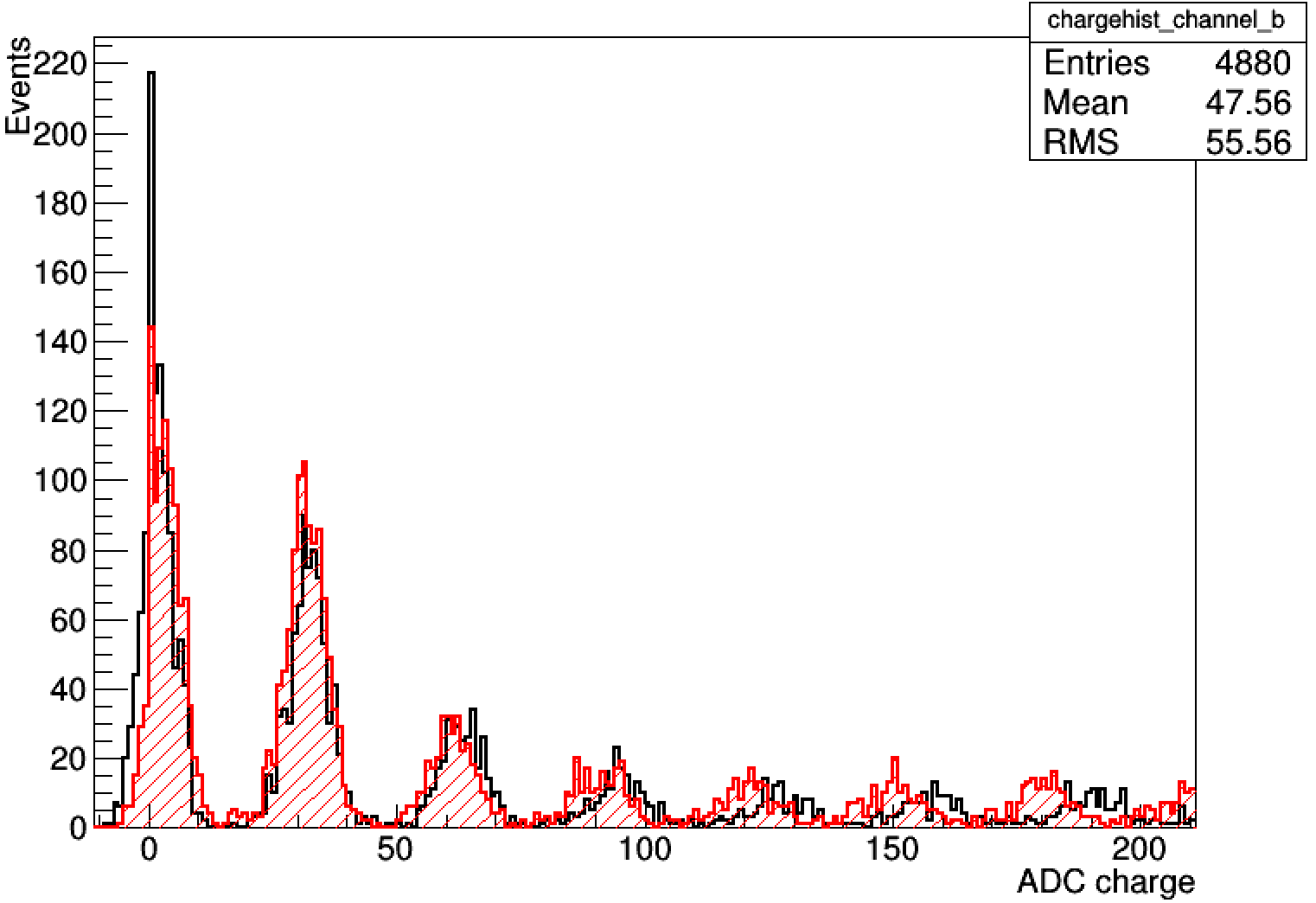}
\caption{Charge spectra for the MPPC S12571-050C, a 50-micron cell size, 1 $\times$ 1 mm$^2$ device. Analogue data from the high gain signal path from the EASIROC chip, digitized with a 12-bit ADC, demonstrates the excellent photo-electron peak-to-peak separation. The EASIROC pre-amp feedback capacitance is set to 100fF, the shaper time constant is set to 50ns. Left) high over voltage leading to $\sim$ 65 ADC/p.e. Right) low over voltage leading to $\sim$ 30 ADC/p.e. Difference in over voltage between left and right acquisitions is 1.75 V.}
\label{50-micron-MPPC}
\end{figure}

\begin{figure}[hbt]
\centering
\includegraphics*[width=0.45\textwidth]{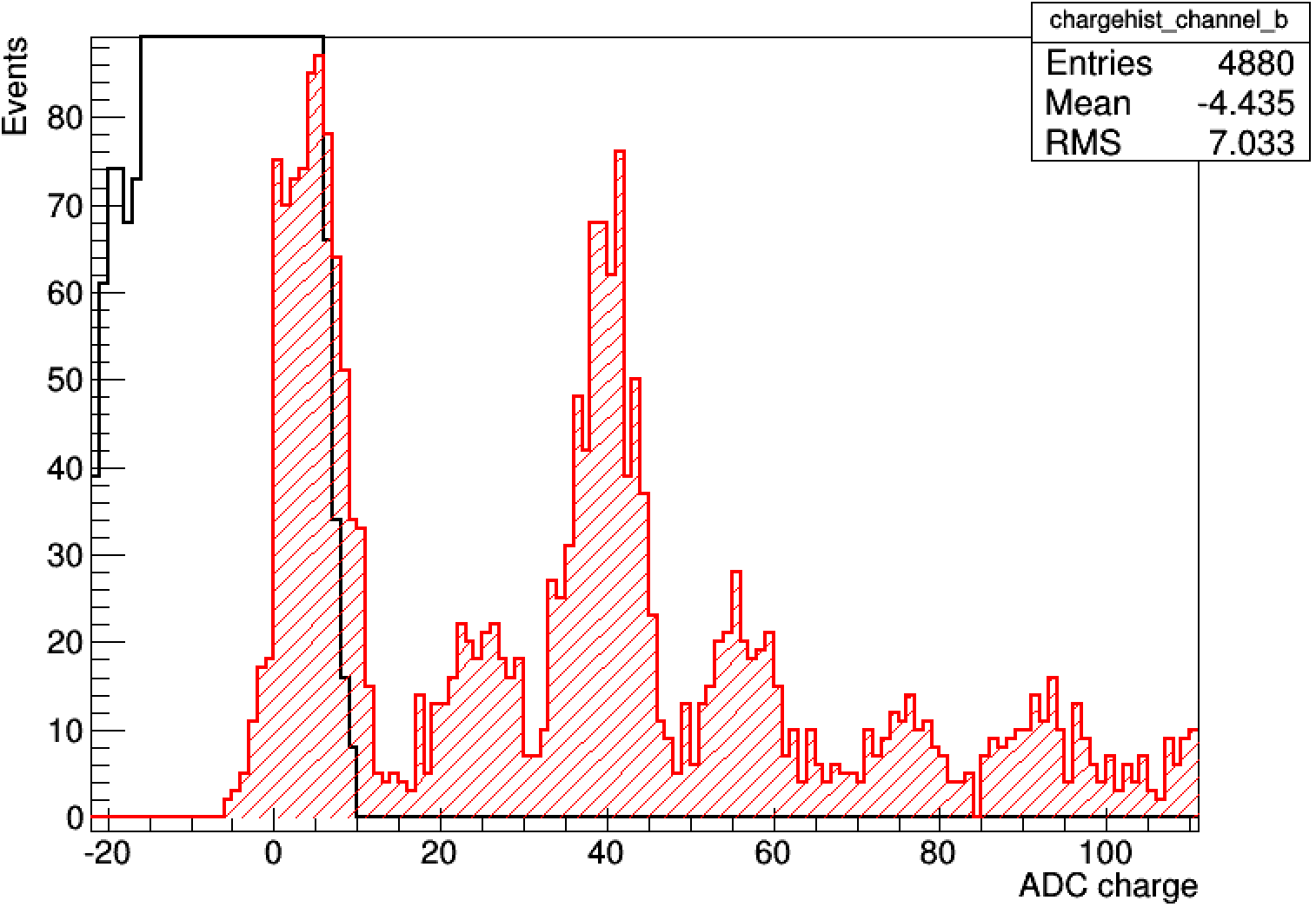}
\includegraphics*[width=0.45\textwidth]{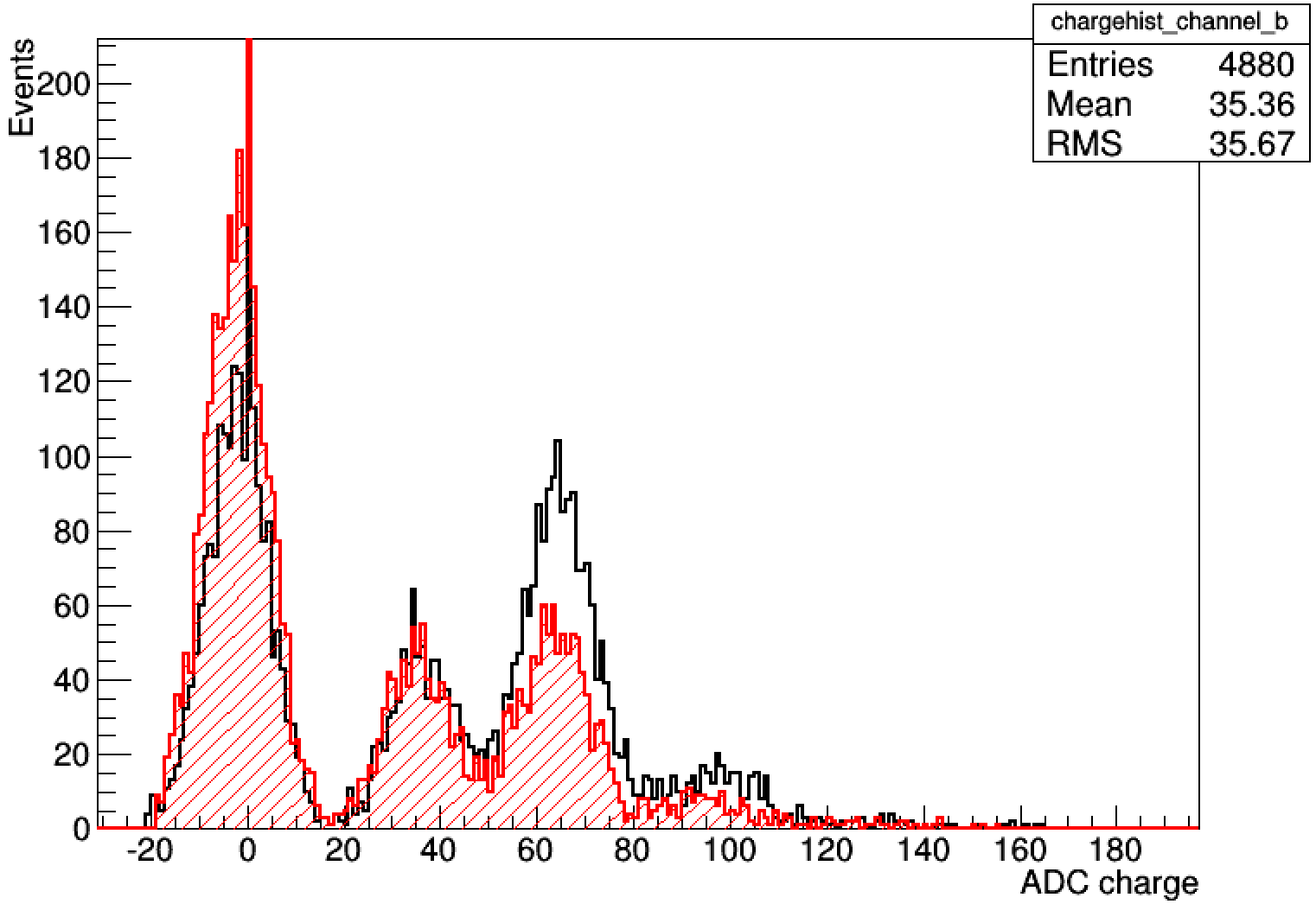}
\caption{Charge spectra for the MPPC S12571-025C, a 25-micron cell size, 1 $\times$ 1 mm$^2$ device. Analogue data from the high gain signal path from the EASIROC chip, digitized with a 12-bit ADC. The EASIROC pre-amp feedback capacitance is set to 100fF for the left plot, and 0fF for the right plot. The shaper time constant is set to 50ns. A relatively low over voltage was used here, it could be increased to provide higher gain.}
\label{25-micron-MPPC}
\end{figure}

The full electronics chain must be taken into consideration before deciding on the photosensor. In order to determine several of the key operational parameters, a test stand was setup with the final configuration of the plastic scintillator bar, a photosensor connector allowing for tests of various photosensors and an evaluation board for the EASIROC chip designed by Omega micro electronics of the IN2P3/CNRS. The test board is described in more detail in another section. Charge information for the photosensors is shown in Figures \ref{50-micron-MPPC} and \ref{25-micron-MPPC}. The exact working point for the photosensors, which can be resumed to the applied overvoltage, is still to be determined from more detailed studies. A large overvoltage leads to a large signal from the photosensor, which in turn leads to high ADC counts per photo-electron once the charge information is digitized by the 12-bit ADCs which are external to the chip. It also leads to a higher signal-to-noise ratio. The disadvantages of a large over voltage are the significantly higher cross-talk, and the smaller electronics dynamic range, which is ultimately limited by the ADC. The electronics dynamic range can be extended to some extent by using the low gain analogue signal path, though it is less straightforward to calibrate, due to the much lower gain, which limits the resolution of single photo-electron peaks.

The following points are therefore relevant in setting the operating point of the photosensors and EASIROC chip, from the perspective of the readout electronics; i.e. the over voltage of the photosensor, and the pre-amp and shaper settings of the EASIROC chip:
\begin{itemize}
	\item{Calibration of signals: peak-to-peak resolution:}
	\begin{itemize}
		\item{should be able to calibrate high gain signal path;}
		\item{should be able to calibrate low gain signal path;}
		\item{this puts stringent requirements on the cross-talk;}
		\item{calibration can be performed at higher pre-amp gain if linear.}
	\end{itemize}
	\item{ADC counts per photo-electron peak:}
	\begin{itemize}
		\item{high enough to resolve individual photo-electron peaks;}
		\item{high enough to provide good signal/noise: noise $\sigma=5 ADC$;}
		\item{required noise level is 0.2 p.e. Pk-to-pk should be $> 25 ADC counts$.}
	\end{itemize}
\end{itemize} 

As an example, setting the chain to obtain 25 ADC/p.e. on the high gain path, with a 12-bit ADC giving a range of 4096 ADC and a baseline around 1000 ADC (could be improved), the full range of the high gain path would be 120 p.e. Using the same assumptions for the low gain signal path but this time with 5 ADC/p.e., the full range of the low gain path would be 620 p.e. This example shows that the system can be optimized for both:
\begin{itemize}
	\item{the most likely events with the high gain signal path (one MIP would yield 100 p.e. for one photosensor, when the event occurs in the plastic close to the photosensor);}
	\item{events with high energy deposition such as stopping particles (estimated at 500 p.e.).}
\end{itemize}

The final choice of photosensor must be made taking the specific application into consideration. Of the commercially available products, two main parameters will drive the choice:
\begin{itemize}
		\item{the size of the photosensor active area, with two options:}
		\begin{itemize}
			\item{1.0 $\times$ 1.0 mm$^2$;}
			\item{1.3 $\times$ 1.3 mm$^2$.}
		\end{itemize}
		\item{the cell size, with two options:}
		\begin{itemize}
			\item{25 $\mu$m cell size;}
			\item{50 $\mu$m cell size.}
		\end{itemize}
\end{itemize}

The size of the photosensor active area must be matched to the light emission profile from the wavelength shifting fiber. The WLS fiber used here is the Y11 double-clad fiber from Kuraray. The total diameter is 1.0 mm, with the double cladding accounting for 6\% of this total, on the outer rim. A light intensity map was obtained using a digital photosensor operating in dark count mode, with a setup representative of the operational environment for the detector modules, reproducing all physical processes occurring from plastic to fiber. A $^{90}$Sr source was used, generating electrons which were incident on the plastic scintillator bars. Light conversion and transfer in the WLS fiber down to the fiber ends was measured, showing that 90\% of the light is transmitted down the fiber core. This result hinted at the possibility of using 1.0 $\times$ 1.0 mm$^2$ devices rather than 1.3 $\times$ 1.3 mm$^2$ devices. The smaller photosensors  are a factor $\times$ 1.5 cheaper. However, their selection requires a good design of the photosensor connector to ensure:
\begin{itemize}
			\item{good alignment of the photosensor with the WLS fiber axis, to within 100 $\mu$m;}
			\item{minimal distance between WLS fiber end and photosensor active area.}
\end{itemize}
Very preliminary results of the latest generation process from Hamamatsu with 50 $\mu$m cell size show the summed light output from both ends of the bar, equipped with the final connector design, to be 127 p.e./MIP for the 1.0 $\times$ 1.0 mm$^2$ devices and 141 p.e./MIP for the 1.3 $\times$ 1.3 mm$^2$ devices.
Concerning the cell size, the factor $\times$ 4 higher dynamic range of the 25 $\mu$m cell size is clearly an advantage. Preliminary tests show that the light yield is not much lower than the 50 $\mu$m cell size devices. However, the lower gain does introduce challenges for calibration when coupled to the EASIROC chip. The peak-to-peak resolution is significantly degraded, hence it is more difficult to distinguish between individual photons.

Hamamatsu are actively working on improved devices with significantly less cross-talk between adjacent pixels which is achieved by surrounding each pixel by a trench. First non-commercial prototypes of these devices will be made available for preliminary testing from June 2014. They introduce the prospect of operating with much higher over voltage, leading to higher gain and PDE. It is therefore expected that the 25 $\mu$m cell size versions of these low cross talk devices (LCT4) will be better matched to the electronics based on the EASIROC chip.

%============================================
\begin{table}[h]
\centering
\caption{\em Comparison of specifications and performance of different photosensors from a range of manufacturers, measured under conditions representative of the detector modules.}
\begin{tabular}{cccccccc}
\toprule
\textbf{Parameter} & \textbf{Unit} & \textbf{MPPC-T2K} &	\textbf{ASD-40}  & 	\textbf{KETEK}  &	\textbf{SensL}\\
\hline
\multicolumn{4}{l}{Manufacturer reported specifications}\\
\hline
Pixel size & ${\mu}m$ & 50	&	40	 &	50	&	20 \\		
Number of pixels	&	& 667	 &	600	&	400 & 848 \\	
Sensitive area	& mm$^2$	& 1.3 $\times$ 1.3	 &	dia 1.2 & 1.0 $\times$ 1.0	&	1.0 $\times$1.0 \\
Gain	& 	& $7.5\times10^5$	 &	$1.6\times10^6$ & -	&	-\\
Dark rate & MHz 	& $\leq1$	 &	$\sim3$ & $\leq2$	&	$\leq2$\\
Bias voltage	& V	& $\sim70$	 &	30-50 & 33-50	&	30\\
\hline
\multicolumn{4}{l}{Performance}\\
\hline
Overvoltage	& V	& $\sim$1.4	 &	3.6 & 4.5	&	2.7\\
Dark rate & kHz 	& 900	 &	3630 & 1250	& 1960\\
Crosstalk & \% 	& 10	 &	13.4 & 35	& 9.7\\
Pulse shape& - 	& good	 &	good & long tails& good\\	
Peak separation& - 	& good	 &	good & bad& bad\\	
PDE& \% 	& 25.6	 &	11 & 26.4& 14.2\\	
\bottomrule
\end{tabular}
\label{photosensorcomparison}
\end{table}
%============================================ 

 %..............................................................................................................................
%..............................................................................................................................  

 %..............................................................................................................................
%.............................................................................................................................. 

\clearpage

\subsection{Scintillator and fiber connectors}
A good geometrical interface between the SiPM sensitive area and the fiber is a crucial step in achieving good signal transmission efficiency and signal quality. Experience gained with the design of the MICE EMR optical connectors at the University of Geneva has proved valuable in designing the photosensor connectors. Connector prototypes were manufactured with 3D lithography. The final connector design and mass production using plastic injection moulding is to be carried out by the INR. Particular attention will be paid to fiber polishing and assembly stages, where quality assurance must be guaranteed and costs and schedule controlled.

\begin{figure}[hbt]
\centering
\includegraphics*[width=0.45\textwidth]{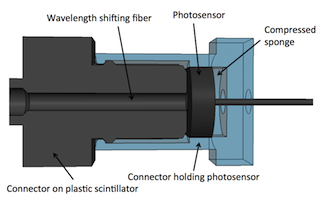}
\hfill
%\includegraphics*[width=0.35\textwidth]{figs/AIDA_V3bis-3.png}
%\hfill
%\includegraphics*[width=0.35\textwidth]{figs/AIDA_V3bis-4.png}
%%\hfill
\includegraphics*[width=0.45\textwidth]{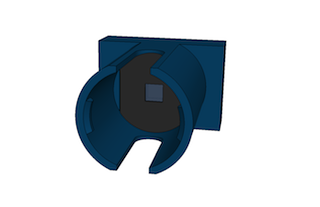}
\hfill
\caption{Photosensor connector concept. The design has evolved from these initial sketches, to include constraints imposed by the plastic injection moulding process.}
\label{connector}
\end{figure}

The concept for a connector system to ensure optimal coupling between the wavelength shifting fibre and photosensor surface is shown in Figure \ref{connector}. It consists of the following main components:
\begin{itemize}
\item The plastic scintillator slab;
\item The wavelength shifting fibre;
\item Connector A, which ensures the WLS fibre is centered with respect to the plastic scintillator slab;
\item The photosensor (MPPC);
\item A component with a spring effect (sponge);
\item Connector B, which holds the photosensor in place;
\item A miniature pcb, which couples to the photosensor pins.
\end{itemize}
The first connector A once glued onto the plastic scintillator bar provides a support enabling the polishing of the fiber ends. The second connector B is designed to house the photosensor. Some parameters for the above components have been fixed, some choices are still possible. The production of plastic scintillator slabs has started, and in this context, the connector concept was designed to allow for production to proceed on the elements that affect the plastic scintillator production (the WLS, connector A).

%..............................................................................................................................
%..............................................................................................................................

The connector concept proposed here is based on a connector for the Hamamatsu MPPC devised for the T2K experiment. It is therefore based on the dimensions of physical objects, such as the MPPC and the pcb for electrical connections which hosts a mini-coaxial connector. Many of the geometrical constraints are very similar, modifications were brought to the design to account for the smaller dimensions of the connector. 

 %..............................................................................................................................
\begin{figure}[hbt]
\centering
\includegraphics*[width=0.45\textwidth]{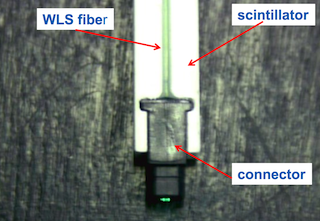}
\includegraphics*[width=0.45\textwidth]{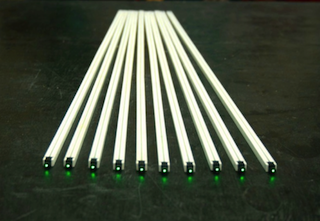}
\caption{Plastic scintillator bars with their connectors, as planned for the neutrino detector prototype modules.}
\end{figure}
%..............................................................................................................................  

\subsection{Detector module mechanics}

\begin{figure}[hbt]
\centering
\includegraphics*[width=0.45\textwidth]{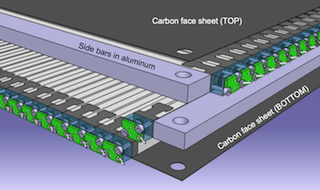}
\hfill
\includegraphics*[width=0.49\textwidth]{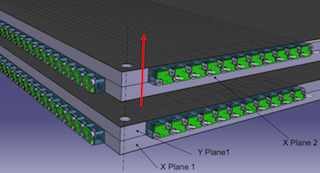}
\caption{TASD and MIND prototype detector module: assembly.}
\label{mechanics}
\end{figure}

The detector module mechanics proposal is shown in Figure \ref{mechanics}. The concept is to have independent modules that can be installed one-by-one either:
\begin{itemize}
	\item{in a mechanical support frame for the TASD inside the Morpurgo magnet;}
	\item{or between steel plates in a MIND.}
\end{itemize}

The assumption is that any repair or maintenance work which might be required on an individual channel can be carried out by sliding the module out of the assembled detector, transporting the module to a workshop, and carrying out the necessary changes to the module. This modularity is driven by our latest experience with the MICE-EMR detector, where the plastic scintillator bars and connectors cannot be worked on unless the whole detector is dismantled.

A module consists of one X plane of 84 scintillator bars glued at either end onto an aluminium support bar and positioned onto one Y plane of 84 scintillator bars also glued at either end onto a second pair of aluminium support bars. The X and Y planes are sandwiched between two carbon or kapton sheets, of thickness $200\, \mathrm{{\mu}m}$. These sheets provide mechanical support, reinforcing the planar alignment of the scintillator bars, and also provide some degree of light tightness. The carbon sheets will not affect the physics capabilities of the MIND detector, since their thickness is negligible with respect to that of the steel plates. For the TASD detector, these sheets add 3\% to the overall thickness of the detector modules. Their impact has not been fully evaluated but is not expected to be significant.

One important issue to be addressed is that of optical crosstalk. It should be determined whether the fiber groove should be covered with a paper reflector in order to limit the amount of light being transmitted from a bar on one plane to bars on another plane.

%..............................................................................................................................
%..............................................................................................................................

\subsection{Electronics and DAQ}
Emphasis will be placed on the electronics options which offer the best opportunity for further development to cover medium-term foreseeable requirements for neutrino detectors and related applications.
The following options have been studied:
\begin{itemize}
\item{DRS4: potential with long term perspectives but currently expensive,}
\item{EASIROC: 3kHz readout rate demonstrated, architecture close to the MICE EMR,}
\item{T2K ND280 TRIP-t option: will not be considered for this test beam.}
\end{itemize}

The data acquisition system will be adapted from the MICE EMR DAQ. The EASIROC readout chip is the baseline solution. R\&D on the DRS4 chip is ongoing at the University of Geneva within the framework of upgrades to the NA61 experiment. Depending on progress, a few modules could be equipped with DRS4 readout boards.

 We plan to adopt a readout system where we readout separately the slower low and high gain analogue signal paths at 1 kHz providing charge information, and the faster hit-only digital triggers representing 4000 samples for every particle trigger running at 400 MHz, covering 10 $\mu$s after each event, with an average readout rate of 100 kHz. The faster digital triggers should allow tagging of delayed signals related to each event (e.g. muon decay to electron). 

\subsubsection{Outline of electronics chain}
The planned electronics chain is based on the electronics chain developed for the MICE-EMR detector, installed on the MICE beamline at the Rutherford Appleton Laboratory in September 2013. It is to be adapted taking into consideration:
\begin{itemize}
	\item{the beam characteristics of the beamline at CERN;}
	\item{the different photosensors, going from PMTs to silicon photomultipliers;}
	\item{the different readout chip, from MAROC to EASIROC.}
\end{itemize}

In particular, a new front-end board must be designed. Coupling of the photosensors to the EASIROC chip can be done using the scheme implemented on the EASIROC evaluation board.

Concepts to be re-used from the MICE-EMR detector include the use of a Digitizer Buffer that acts as a TDC, assigning a time stamp to every event. The VME Readout Board (VRB) will also be re-used. It collects information from the front-end memory buffers and transmits it to the PC.

 %..............................................................................................................................
\begin{figure}[hbt]
\centering
\includegraphics*[width=0.7\textwidth]{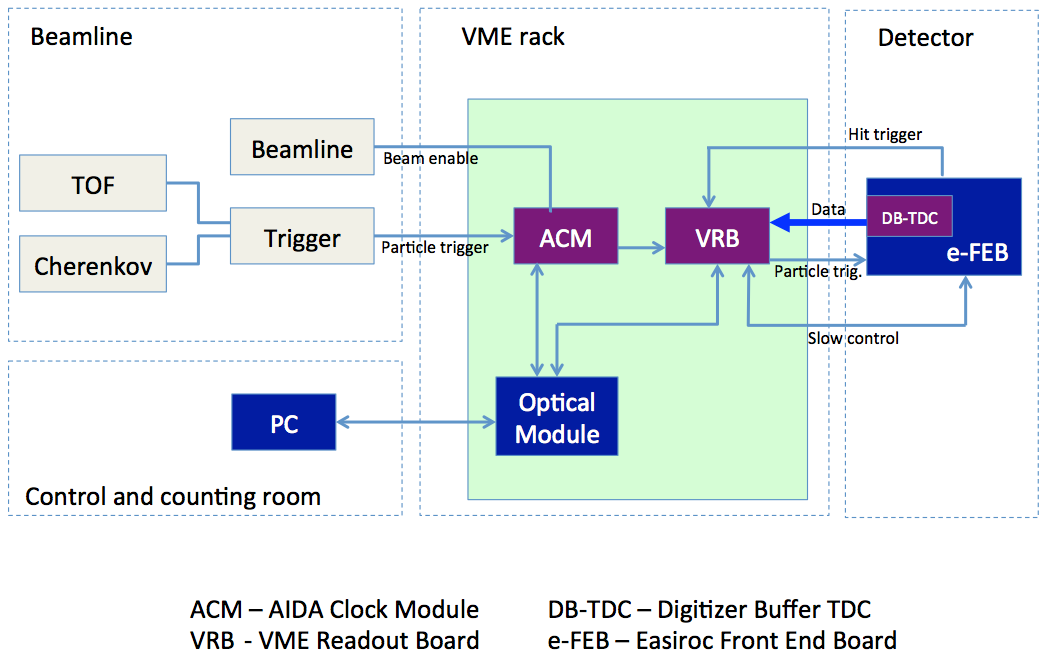}
\caption{Overview of the electronics chain for the neutrino detector prototypes.}
\label{electronics-scheme}
\end{figure}
%.............................................................................................................................. 

\subsubsection{Beam considerations for electronics}
The planned set of runs in the North Area foresees the implementation of a very low energy beam line, delivering protons, pions, electrons, muons in the energy range 0.5 to 9 GeV/c, Table \ref{particlerates}. The beam is a slow extracted beam with a pulse length of up to 10 s, delivering particles at a rate of 1 kHz or so, Figure \ref{spsslowextraction}. The pulse repetition frequency is maximum 0.03 Hz. 
 
 %..............................................................................................................................
\begin{figure}[hbt]
\centering
\includegraphics*[width=0.8\textwidth]{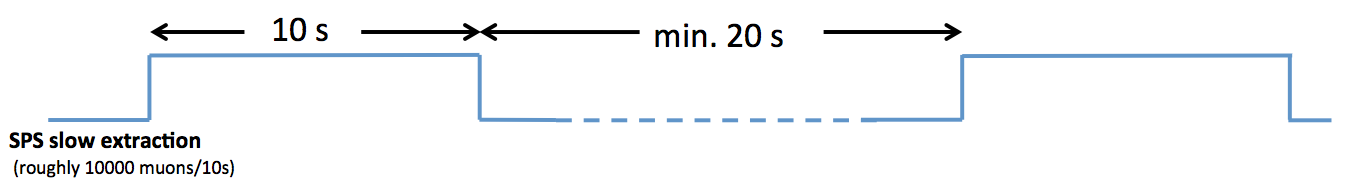}
\caption{SPS slow extraction beam time structure. These assumptions are taken in the design of the neutrino detector electronics.}
\label{spsslowextraction}
\end{figure}
%..............................................................................................................................
 
 %..............................................................................................................................
\begin{figure}[hbt]
\centering
\includegraphics*[width=0.8\textwidth]{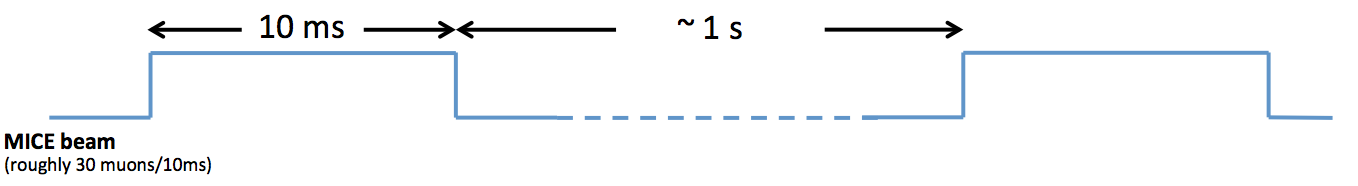}
\caption{MICE beam structure. The downstream components of the MICE electronics chain will be adapted for use in the readout electronics chain.}
\label{spsslowextraction}
\end{figure}
%.............................................................................................................................. 

 %..............................................................................................................................
\begin{figure}[hbt]
\centering
\includegraphics*[width=0.8\textwidth]{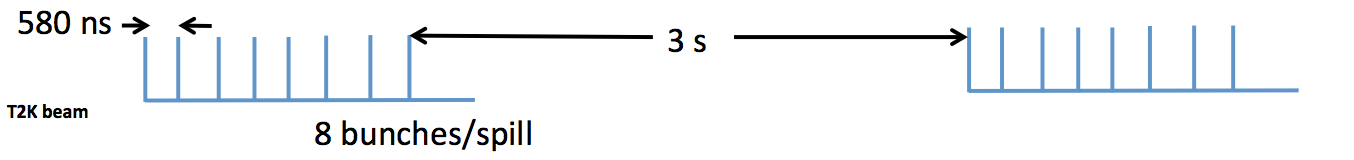}
\caption{T2K beam structure. The Trip-t electronics for T2K were initially chosen as the baseline but are no longer being considered.}
\label{spsslowextraction}
\end{figure}
%.............................................................................................................................. 

The background is expected to be mainly from high energy muons created from the interaction of the primary beam with the first target. It should be low enough that triggering on interesting events will be achievable with high efficiencies.

\subsubsection{EASIROC chip}
The EASIROC (Extended Analogue SI-pm ReadOut Chip) chip is designed in 0.35 $\mu$m SiGe technology, with first versions available since 2010 and used in a variety of experiments such as PEBS, MuRAY, at JPARC and in medical imaging. It is a 32 channel fully analogue front-end ASIC dedicated to readout of SiPM photosensors. It is derived from the SPIROC chip which was developed for hadron calorimetry foreseen for the International Linear Collider.

The chip integrates a 4.5 V (2.5 V) range 8-bit DAC for individual SiPM gain adjustment. A multiplexed charge measurement is available from 160 fC to 320 pC with 2 analogue outputs. These charge paths are made of 2 variable gain preamplifiers followed by 2 tunable shapers and a track and hold. 

The analogue core is sensitive to positive SiPM signals. For each channel, two parallel AC coupled voltage preamplifiers ensure the read out of the charge from 160 fC to 320 pC (ie. 1 to 2000 photoelectrons with SiPM Gain = $10^6$, with a photoelectron to noise ratio of 10). Two variable shapers are used to reduce noise; each of them has an adjustable peaking time from 25 to 175 ns to allow the user to minimize the noise depending on the final application. A trigger line is available from the high gain preamplifier. It is composed of a 15 ns peaking time fast shaper followed by a discriminator. The threshold is set by an internal 10-bit DAC and is common to the 32 channels. The 32 triggers are available as a 32-bit output bus and can be either latched or directly outputted.

In addition to the charge output, timing measurements are possible via a trigger path, that integrates a fast shaper followed by a discriminator. It's threshold can be set via a common 10-bit DAC. The 32 trigger outputs are complemented by an OR32 output. 

Power consumption is 5 mW/channel and unused features can be powered OFF. 

%..............................................................................................................................
\begin{figure}[hbt]
\centering
\includegraphics*[width=0.30\textwidth]{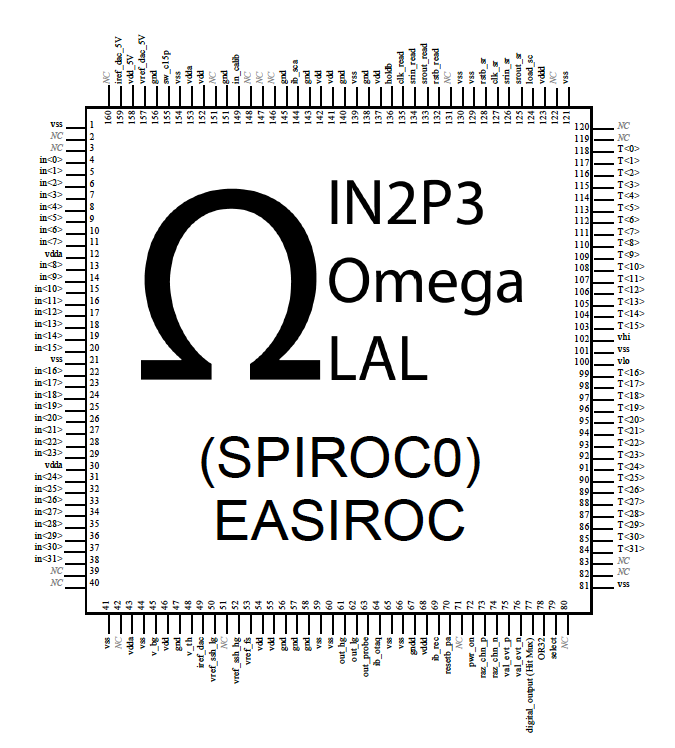}
\includegraphics*[width=0.55\textwidth]{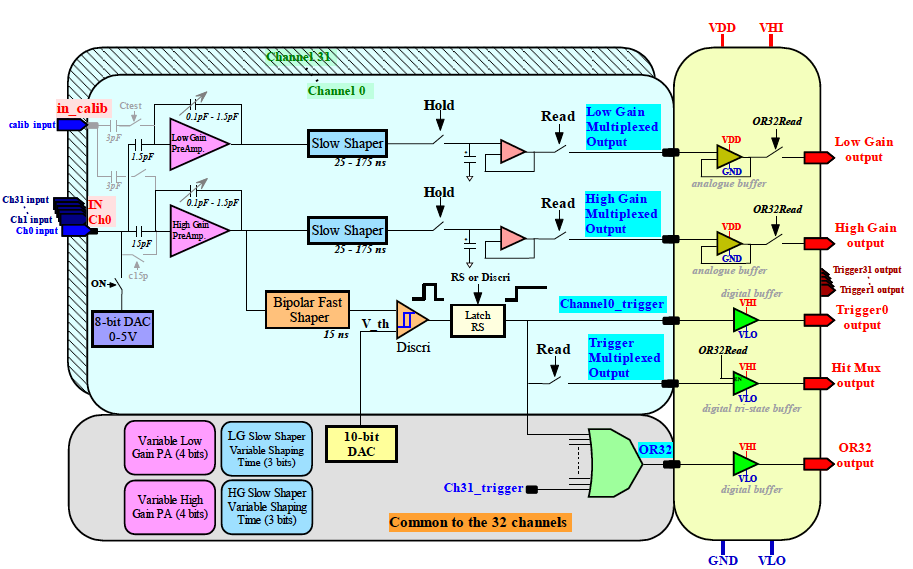}
\caption{Pinout and schematic of the EASIROC chip.}
\label{easiroc-pins}
\end{figure}
%.............................................................................................................................. 

\subsubsection{Readout modes}
As described above, two readout modes from the EASIROC are possible:
\begin{itemize}
	\item{Mode A: slow readout of the analogue multiplexed signals;}
	\item{Mode B: Fast readout of the 32 triggers.}
\end{itemize}

\begin{figure}[hbt]
\centering
\includegraphics*[width=0.6\textwidth]{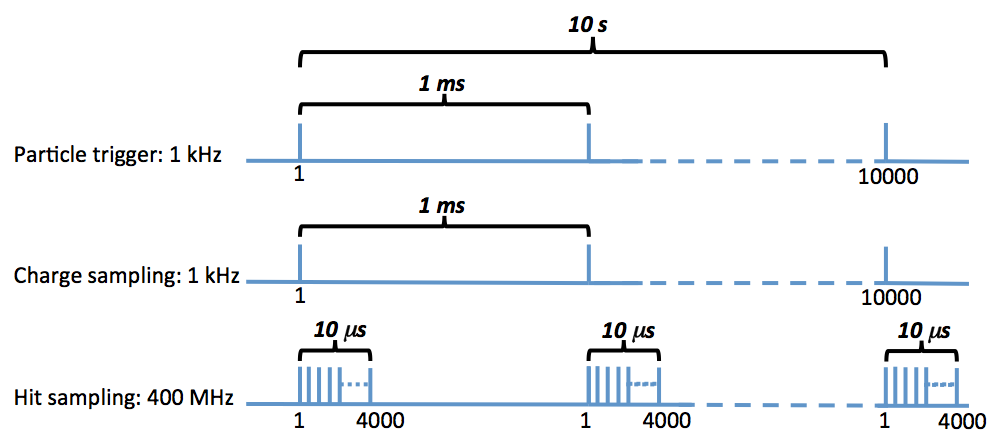}
\caption{Particle triggers, charge sampling and hit sampling within a 10s slow extraction spill from the SPS.}
\label{sampling-rates}
\end{figure}

Used in combination, it is planned to operate with an external particle trigger provided by a system upstream, for example a Time of Flight detector, at rates of few $\sim$kHz. Mode A would then provide one charge sample per particle trigger. Mode B can be operated at 400 MHz, recording 10 $\mu$s-worth of digital hits occurring immediately after the primary event, such as decay products, i.e. 4000 samples per particle trigger, see Figure \ref{sampling-rates}.

%..............................................................................................................................
%.............................................................................................................................. 

The Front-End Board will be designed with sufficient memory buffers to record all data within a spill, lasting 10s in the case of a slow extracted beam at the SPS. The stored data will be transmitted between spills to the VME Readout Board (VRB), and then onto a PC. The minimum length between spills is 20s.

\subsubsection{Data stream}
 It is planned to have one Front End Board per plane, i.e. 90 ch/FEB. However, detailed studies are required, including costing, before committing to a 1 FEB/plane scheme. Another possibility is to have one FEB/EASIROC chip, i.e. 30 ch/FEB, 3 FEB/plane. Since only roughly 1/3 of channels will be instrumented in phase 1, it is to be decided whether the corresponding fraction of each module will be instrumented, or whether full modules will be instrumented, but only 1/3 of the total number of modules will be used.

%..............................................................................................................................
\begin{figure}[hbt]
\centering
\includegraphics*[width=0.7\textwidth]{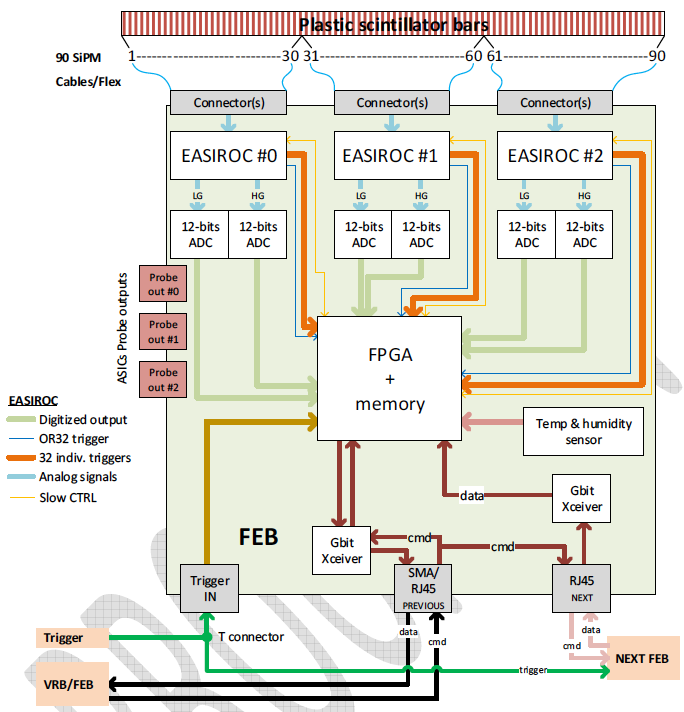}
\caption{Readout scheme for one plane.}
\label{Electronics-sketch}
\end{figure}
%.............................................................................................................................. 

The bit allocation scheme is reported in Table \ref{bit-allocation}. The scheme was devised to allow flexibility in the recording of time, with several configurations possible by combining the 16-bit \mbox{\bf{Hit time measurement}} with the 20-bit \mbox{\bf{Event time tag}}. 

For timing under normal operation at the SPS with a slow extraction spill, the hit time measurement should cover 10 $\mu$s with sampling every 2.5 ns: 12 bits are sufficient.
The event time tag should reset every 10 $\mu$s, a rate of 100 kHz, and cover 10s which is the spill length: 20 bits are required.
%============================================
\begin{table}[h]
\centering
\caption{\em Allocated bits for different parameters in the recorded data.}
\begin{tabular}{cccccccc}
\toprule
\textbf{Parameter} & \textbf{Allocated bits} & \textbf{Full range} &	\textbf{Requirement} 	\\
\hline
Word type & 4 & 16	&	-	 \\
Board ID & 10 & 1024	&	100	 \\
Spill number & 16 & 65536	&	-	 \\
Event count (Ext. particle trigger) & 28 & 268435456	&	1e8	 \\
Event ID (TBC) & - & -	&	-	 \\
Channel ID & 7 & 128	&	90	 \\
Hit ID & 5 & 32	&	-	 \\
Hit time measurement & 16 & 65536	 &	4000	 \\
Amplitude status & 4 & 16	&	2	 \\
Hit amplitude measurement & 12 & 4096	&	4096	 \\
Hit count within event & 6 & 64	&	-	 \\
Event time tag & 20 & 1048576	&	1e6	 \\
Spill width (TBC) & 22 & 4194304	&	-	 \\
\bottomrule
\end{tabular}
\label{bit-allocation}
\end{table}
%============================================ 
% 
%  %..............................................................................................................................
%\begin{figure}[hbt]
%\centering
%\includegraphics*[width=1.0\textwidth]{figs/AIDA-data-structure.png}
%\caption{Data structure for the neutrino detector prototypes.}
%\label{detectormodules}
%\end{figure}
%%.............................................................................................................................. 

The estimation of data rate is critical in defining the buffer size for memory on the Front End Board. Several assumptions are made, these are usually conservative. Experience with the MICE detector would suggest that during normal data taking, the buffers are over-dimensioned by a large factor. However, some events, such as electromagnetic or hadronic showers, lead to a large number of hits, which must be recorded.

Table \ref{data-rates} lists the expected data rates for runs under the assumptions of a slow extracted beam at the SPS outlined previously. In summary, we expect \mbox{\bf{1 MB/plane/spill}}, where a spill is 10s. The VRB memory buffer of the existing MICE-EMR detector is 8 MB. With one FEB/plane, it is relatively straightforward to reconfigure the firmware of the VRBs so that each VRB serves 8 planes, or 4 modules.
A fraction of the 9000 plastic scintillator bars will be instrumented in the first phase. 3000 channels will be instrumented, we would therefore need 34 FEBs, 5 VRBs, 1 VME crate. The transfer rate from VRB to PC was measured to be 80 $\mu$s per kB, it should be re-measured with the configuration described here. It would lead to transfer times of 2.5s/event for the instrumented part of the detector, i.e. 3000 channels.

%============================================
\begin{table}[h]
\centering
\caption{\em Data rates estimated during operation of the neutrino detector prototypes. Note that in the first phase, only a fraction of the whole detector, 3000 of 9000 bars, will be instrumented.}
\begin{tabular}{cccccccc}
\toprule
\textbf{Parameter}  & 	\textbf{Per plane}  &	\textbf{Instrumented detector (Whole)}\\
\hline
\multicolumn{3}{l}{Items per plane}\\
\hline
\# channels		 &	90	&	3000 (9000) \\		
\# FEB	 &	1	&	34 (100) 	 \\
\# EASIROC	 &	3	&	102 (300) 	 \\
\hline
\multicolumn{3}{l}{Charge and hits per particle trigger(event)}\\
\hline			
\# Charge output/Particle trig/plane &	10	&	-  \\
\# Hit output/Particle trig/plane &	10	&	-  \\
\hline
\multicolumn{3}{l}{Stored information}\\
\hline	
Stored Bytes/event	&	92.0	&	 \\		
Particle triggers/spill	 	&	10000 &	 \\
Stored Bytes/spill [MB]	&	0.920 & 31.28	 \\			
\bottomrule
\end{tabular}
\label{data-rates}
\end{table}
%============================================ 

\subsubsection{Slow control}
It is planned to implement the slow control via the VRB chain, i.e. one link for both data and control.

\clearpage
 
\section{Beam request}
\begin{figure}[hbt]
\centering
\includegraphics*[width=1.0\textwidth]{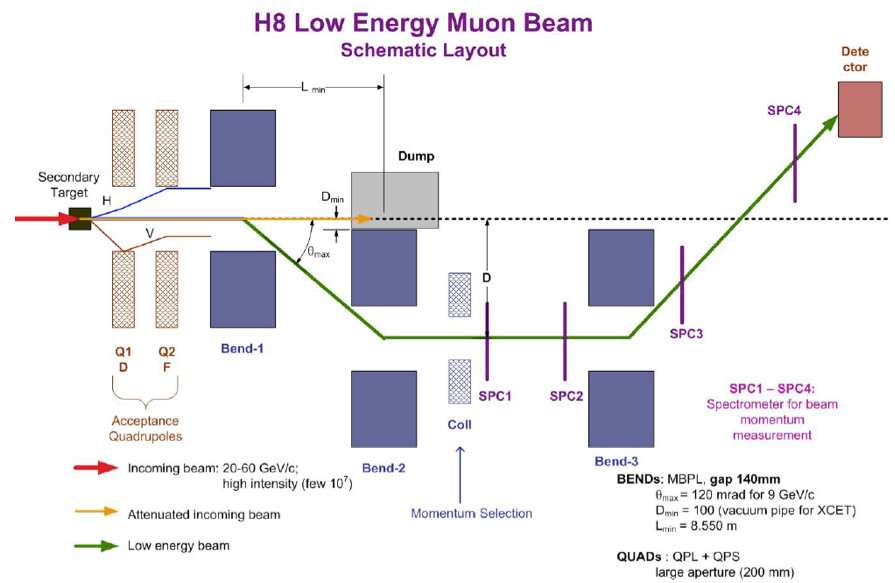}
\caption{Possible implementation of a low energy muon test beam. Note that since it is planned to have the MIND detector located downstream of the TASD detector (assuming the TASD is installed in the MORPURGO magnet), the particle beam must be parallel to the axes of both the TASD and the MIND}
\label{beamline}
\end{figure}

%\mbox{\bf{Include reference to AIDA WP8.2.1 document}}
The test beam is required to deliver electrons, muons and hadrons (pions and protons) in a momentum range between 0.5 and 5.0 GeV/c with the possibility of extending up to 9 GeV/c. A large aperture magnet such as the MORPURGO magnet installed in the North Area at CERN should be included in the test beam infrastructure. A possible location is the H8 beamline at the North Area (which includes the MORPURGO magnet) but could also be in the East Area which would require the installation of a suitable magnet. These options were studied within the European FP7 project AIDA under Work Package 8.2.
A possible implementation is shown in Figure \ref{beamline}. The bend is required because the proton beam impinging on the secondary target will create high energy muons ~40-100 GeV, beam size 1 $\times$ 1 m$^2$, which would be a significant background for the detectors. If the bend is not implemented, a high resolution ~10 ps, large area (1 $\times$ 1 m$^2$) time of flight detector would be required just after the secondary target to veto the high energy muons.
Close collaboration between detector studies and beam studies is required because a number of parameters such as particle rate, beam size and angle of incidence on the detector will affect the detector design.

%..............................................................................................................................
%..............................................................................................................................

\subsection{Particle rates}
Particle rates presented in Table \ref{particlerates} are given as an indication, in order to have a rough idea of the estimated beam time for sufficient statistics, and for an upper limit to be considered in the design of the electronics and data acquisition chains. Significant departures from these values are expected when studies of the beamline in WP8.2 are carried out, especially for low momenta particles. The rate assumed is 1 kHz, with a sample of $10^7$ particles for each energy bin and configuration. It is recognized that a sample of $10^6$ is probably adequate, in which case we can accept a drop in rate to 100 Hz, whilst keeping the same overall test beam time online, estimated to be 16 weeks if we make the assumption of 50 $\times$ $10^8$ triggers in total at 1 kHz with 50\% running efficiency.

%============================================
\begin{table}[h]
\centering
\caption{\em Requirements for particles and their momenta. The particle rate here is the rate within a spill, regardless of the spill length, slow extraction is assumed.}
\begin{tabular}{cccccccc}
\toprule
\textbf{Type} &	\textbf{Momentum [GeV/c]}  & 	\textbf{Rate [kHz]}  &	\textbf{Total} & \textbf{Time est. [hrs]}\\
\hline
\multicolumn{5}{l}{Electron and muon charge separation: TASD in large aperture magnet}\\
\hline
$e^{+/-}$	&	0.5, 0.7, 1.0, 2.0, 5.0, (9.0)	 &	1.0	&	$10^7$$\times$10 &	170 \\		
$\mu$$^{+/-}$	&	0.5, 0.7, 1.0, 2.0, 5.0, (9.0)	 &	1.0	&	$10^7$$\times$10 &	170 \\
\hline
\multicolumn{5}{l}{muon charge separation: MIND}\\
\hline			
$\mu$$^{+/-}$&	0.8, 1.0, 1.5, 2.0, 5.0	(9.0) &	1.0	&	$10^7$$\times$10 &	170 \\
\hline
\multicolumn{5}{l}{hadronic shower reconstruction: MIND}\\
\hline	
$\pi$$^{+/-}$	&	0.5, 0.7, 1.0, 2.0, 5.0, (9.0)	 &	1.0	&	$10^7$$\times$12 &	200 \\		
$p$	&	0.5, 0.7, 1.0, 2.0, 5.0, (9.0)	 &	1.0	&	$10^7$$\times$6 &	100 \\			
\bottomrule
\end{tabular}
\label{particlerates}
\end{table}
%============================================

\subsection{Beam composition}
\begin{figure}[hbt]
\centering
\includegraphics*[width=1.0\textwidth]{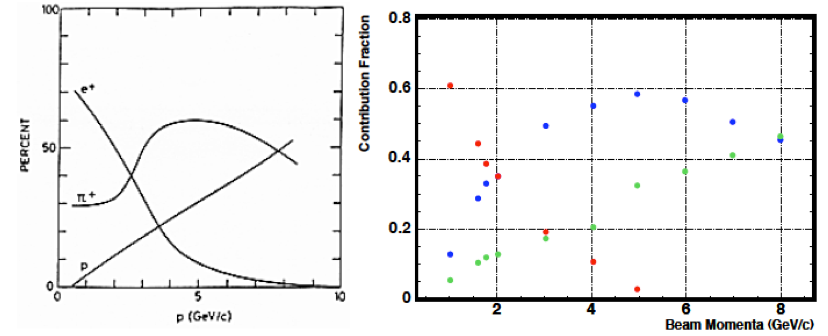}
\caption{PS T7 beamline particle composition. The expected composition is shown on the left. The measured composition is shown on the right [Ref. Anatael thesis].}
\label{beamcomposition}
\end{figure}

The beam composition will be determined from simulations of the beamline. An example is given here of the beam composition from the PS T7 during the MINOS CalDet tests, Figure \ref{beamcomposition}. Similar data are required to know the beam contamination for each of the required particle types in Table 3. Knowledge of the momentum spread for the beam that will be received is required in order to determine whether pion/muon separation at low momenta can be done in the TASD by range.
%..............................................................................................................................
%..............................................................................................................................

\subsection{Large aperture magnet: Morpurgo}
\begin{figure}[hbt]
\centering
\includegraphics*[width=0.35\textwidth]{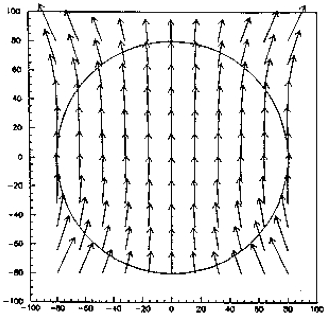}
\hfill
\includegraphics*[width=0.55\textwidth]{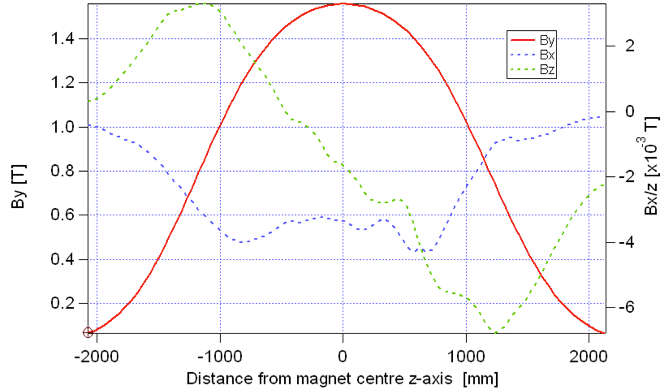}
\caption{Field maps of the Morpurgo magnet: a) the XY plane shows a good approximation to a dipole field, b) the field as a function of Z shows a large variation with respect to the peak value.}
\label{morpurgofield}
\end{figure}

The MORPURGO magnet is a good approximation to a dipole as can be seen in Figure \ref{morpurgofield}, showing the XY plane. Some optimisation of the TASD detector positioning along the z-axis will be required since the peak field value drops sharply from the center of the magnet to its edges, Figure \ref{morpurgofield}.

%..............................................................................................................................
%..............................................................................................................................

\subsection{Test beam programme}
In anticipation of this test beam programme, the Morpurgo magnet would need to be powered and trained for a duration of 1-3 months before the start of the online data taking.
We plan to proceed with the TASD tests first, using the Morpurgo magnet. When the TASD tests are completed, the Morpurgo magnet must be switched off in order to proceed with tests of the MIND detector, since the charged particle beam will travel through the Morpurgo magnet axis to the MIND detector located downstream. We require a period of 2 weeks between the end of runs with the TASD in the Morpurgo magnet and subsequent runs with the MIND.

%\mbox{W$\cdot$m${}^{-3}$}

\clearpage

\subsection{Possible detector layout at H8}
\begin{figure}[hbt]
\centering
\includegraphics*[width=0.8\textwidth]{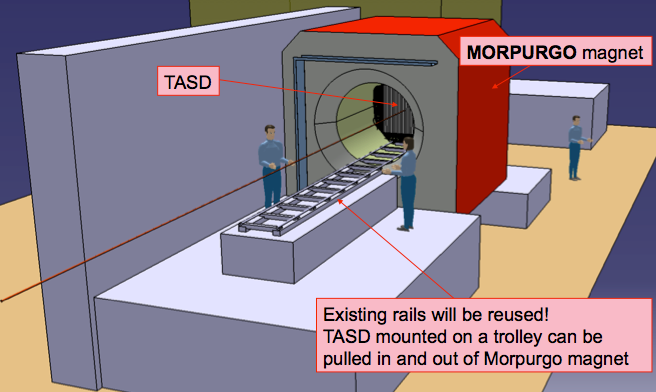}
\caption{Installation of the TASD inside the Morpurgo magnet volume in H8.}
\label{tasdlayout}
\end{figure}

\begin{figure}[hbt]
\centering
\includegraphics*[width=0.8\textwidth]{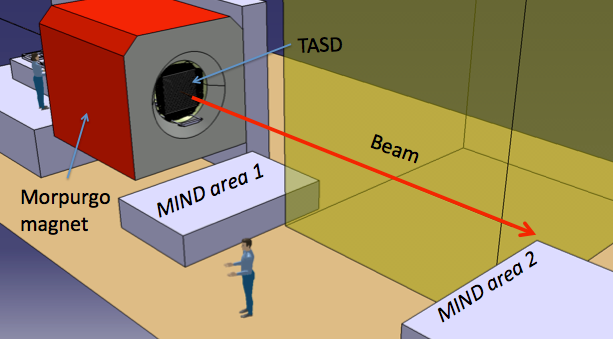}
\caption{Two options to position the baby MIND detector: MIND area 1 could be used for initial tests, although the close proximity to the Morpurgo magnet is an issue to be investigated; MIND area 2 is the proposed final position of the detector.}
\label{mindprotolayout1}
\end{figure}

Both TASD and baby MIND prototypes are designed with the assumption that they will be installed on the H8 beam line, in zone 158 at the North Area. They will be operated separately, which will allow for the re-use of the scintillator modules, hence cost savings. It is planned to install the TASD prototype inside the Morpurgo magnet, and the baby MIND downstream of the TASD in one of two positions as shown in Figures \ref{tasdlayout} \& \ref{mindprotolayout1}. 

%..............................................................................................................................
%..............................................................................................................................

%..............................................................................................................................
%..............................................................................................................................

\clearpage

\clearpage

%\clearpage

\section*{Appendix A: baby MIND prototype parameters}
%============================================
\begin{table}[h]
\centering
\caption{\em Baby MIND prototype parameters.}
\begin{tabular}{lccccccc}
\toprule
\textbf{Parameter} &	\textbf{Symbol}  & 	\textbf{Unit}  &	\textbf{Nominal value} & \textbf{Range Min.} &	\textbf{Range Max.}\\
\hline
\multicolumn{5}{l}{Detector global dimensions}\\
\hline
Detector width	&	$w_{det}$	&	m	&	3.0	&	2.5	&	4.0 \\	
Detector height	&	$h_{det}$	&	m	&	1.5	&	1.0	&	2.0 \\	
Detector depth	&	$d_{det}$	&	m	&	2.3	&	2.0	&	3.0 \\	
\hline
\multicolumn{5}{l}{Iron plates}\\
\hline
Material grade &	- &	-&	AISI1010 &	-	&	ARMCO \\
Number of plates &	$n_{iron}$ &	-	&	51 &	-	&	- \\
Iron width &	$w_{iron}$ &	m	&	3.0 &	2.5	&	4.0 \\
Iron height &	$h_{iron}$ &	m	&	1.5 &	1.0	&	2.0 \\
Iron thickness &	$t_{iron}$ &	cm	&	3.0 &	1.0	&	5.0 \\
Total iron mass &	$m_{iron}$	&	tons	&	54 &	-	&	-	\\
Total iron area &	$a_{iron}$ &	m2	&	230 &	-	&	- \\
Number of slots for coil &	$n_{slots}$ &	-	&	2 &	2	&	4 \\
Slot for coil: width &	$w_{slot}$ &	cm	&	30.0 &	10.0	&	40.0	\\
Slot for coil: height &	$h_{slot}$ &	cm	&	30.0 &	10.0	&	40.0 \\
Support structure &	- &	-	&	TBD &	-	&	- \\
\hline
\multicolumn{5}{l}{Gap between iron plates}\\
\hline
Number of gaps	&	$n_{gaps}$	&	-	&	50	&	-	&	- \\	
Gap thickness	&	$t_{gap}$	&	cm	&	1.65	&	1.6	&	1.7  \\	
Material	&	-	&	-	&	air + plastic	&	-	&	- \\	
\hline
\multicolumn{5}{l}{Plastic scintillator}\\
\hline
Material &	- &	-&	Polysterene &	-	&	-	& \\
Number of modules (xy or uv)	&	$n_{module,tot}$ &	-	&	52.0 &	-	&	- \\
Number of planes per module	&	$n_{plane,mod}$ &	-	&	2.0 &	-	&	- \\
Gap between planes within module &	- &	cm	&	0 &	0	&	0.05 \\
Module envelope thickness &	$t_{env}$	&	cm	&	0.02 &	0	&	0.05	\\
Module total thickness (incl. glue) &	$t_{mod}$	&	cm	&	1.6 & 	1.5	&	1.7	\\
Scintillator bar length &	$l_{sci}$ &	cm	&	90.0 &	80.0	&	100.0 \\
Scintillator bar width &	$w_{sci}$ &	cm	&	1.0 &	0.95	&	1.05 \\
Scintillator bar height &	$h_{sci}$ &	cm	&	0.725 &	0.7	&	0.75	\\
Bars per plane &	$n_{bars,pla}$ &	-	&	84 &	-	&	- \\
Bars per module &	$n_{bars,mod}$ &	-	&	168 &	-	&	- \\
Total number of bars &	$n_{bars,tot}$ &	-	&	8736 &	-	&	-	\\			
\hline
\multicolumn{5}{l}{Light readout and conversion}\\
\hline	
Light readout optical fibres &	- &	-	&	WLS &	C.F.	& -	\\	
Total length of fibre &	fibre &	m	&	12000 &	10000	&	20000 \\
Readout device &- &	-	&	SiPM &	- &	-	\\
Readout channels per bar &- &	-	&	1 &	1&	2	\\
\bottomrule
\end{tabular}
\label{mindparameters}
\end{table}

\clearpage
%============================================

\section*{Appendix B: TASD prototype parameters}
%============================================
\begin{table}[h]
\centering
\caption{\em TASD prototype parameters.}
\begin{tabular}{lccccccc}
\toprule
\textbf{Parameter} &	\textbf{Symbol}  & 	\textbf{Unit}  &	\textbf{Nominal val.} & \textbf{Min.} &	\textbf{Max.}\\
\hline
\multicolumn{5}{l}{Detector global dimensions}\\
\hline
Detector width	&	$w_{det}$	&	m	&	1.0	&	0.9	&	1.1 \\	
Detector height	&	$h_{det}$	&	m	&	1.0	&	0.9	&	1.1 \\	
Detector depth	&	$d_{det}$	&	m	&	0.86	&	-	&	- \\		
Detector depth with gaps &	$d_{gap}$ &	m	&	213.0 &	-	&	- \\
\hline
\multicolumn{5}{l}{Plastic scintillator}\\
\hline
Material &	- &	-&	Polysterene &	-	&	-	& \\
Number of modules (xy or uv)	&	$n_{module,tot}$ &	-	&	52.0 &	-	&	- \\
Number of planes per module	&	$n_{plane,mod}$ &	-	&	2.0 &	-	&	- \\
Gap between planes within module &	- &	cm	&	0 &	0	&	0.05 \\
Module envelope thickness &	$t_{env}$	&	cm	&	0.02 &	0	&	0.05	\\
Module total thickness (incl. glue)&	$t_{mod}$	&	cm	&	1.6 & 	1.5	&	1.7	\\
Scintillator bar length &	$l_{sci}$ &	cm	&	90.0 &	80.0	&	100.0 \\
Scintillator bar width &	$w_{sci}$ &	cm	&	1.0 &	0.95	&	1.05 \\
Scintillator bar height &	$h_{sci}$ &	cm	&	0.725 &	0.7	&	0.75	\\
Bars per plane &	$n_{bars,pla}$ &	-	&	84 &	-	&	- \\
Bars per module &	$n_{bars,mod}$ &	-	&	168 &	-	&	- \\
Total number of bars &	$n_{bars,tot}$ &	-	&	8736 &	-	&	-	\\			
\hline
\multicolumn{5}{l}{Gap between scintillator modules}\\
\hline	
Number of gaps &	$n_{gaps}$ &	-	&	51 &	-	&	-	\\	
Gap thickness &	$t_{gap}$ &	cm	&	2.5 &	0.0	&	2.5 \\
Material &- &	-	&	air &	vac.	&	tungsten	\\
\hline
\multicolumn{5}{l}{Light readout and conversion}\\
\hline	
Light readout optical fibres &	- &	-	&	WLS &	C.F.	& -	\\	
Total length of fibre &	$l_{fibre}$ &	m	&	12000 &	10000	&	20000 \\
Readout device &- &	-	&	SiPM &	- &	-	\\
Readout channels per bar &- &	-	&	1 &	1&	2	\\
\bottomrule
\end{tabular}
\label{tasdparameters}
\end{table}
%============================================

%\clearpage

\clearpage

%\bibliographystyle{plain}
%\bibliography{sps_bib}

\end{document}